%% file: ms.tex
\documentclass[12pt]{spieman}

\usepackage{amsmath,amsfonts,amssymb}
\usepackage{graphicx}  
\usepackage{setspace}
\usepackage{tocloft}

\usepackage{deluxetable}  

\def\wfirst{\textit {Roman}}

\def\Lstar{L_{\star}}
\def\REarth{r_\oplus}
\def\MEarth{M_\oplus}
\def\nEarth{\eta_\oplus}
\def\MSun{M_\odot}

\def\LSun{L_\odot}

\def\RJup{r_{\rm Jup}}
\def\gapp{\lower 3pt\hbox{${\buildrel > \over \sim}$}\ }
\def\lapp{\lower 3pt\hbox{${\buildrel < \over \sim}$}\ }
\def\proptosim{\lower 3pt\hbox{${\buildrel \propto \over \sim}$}\ }

\def\arcsec{$^{\prime\prime}$}
\def\degree{$^\circ$}

\newcommand\farcs{\mbox{$.\!\!^{\prime\prime}$}}

\title{Starshade Rendezvous: \\
  Exoplanet Sensitivity and Observing Strategy}

\author[a]{Andrew~Romero-Wolf}
\author[a]{Geoffrey~Bryden}
\author[b]{Sara Seager}
\author[c]{N. Jeremy Kasdin}
\author[a]{Jeff Booth}
\author[d]{Matt Greenhouse}
\author[a]{Doug Lisman}
\author[e]{Bruce Macintosh}
\author[a]{Stuart Shaklan}
\author[d]{Melissa Vess}
\author[f]{Steve Warwick}
\author[a]{David Webb}
\author[a]{John Ziemer}
\author[a]{Andrew Gray}
\author[a]{Michael Hughes}
\author[a]{Greg Agnes}
\author[f]{Jonathan W. Arenberg}
\author[a]{S. Case Bradford}
\author[a]{Michael Fong}
\author[a]{Jennifer Gregory}
\author[a]{Steve Matousek}
\author[a]{Jason Rhodes}
\author[a]{Phil Willems}
\author[g]{Simone D'Amico}
\author[h]{John Debes}
\author[d]{Shawn Domagal-Goldman}
\author[a]{Sergi Hildebrandt}
\author[a]{Renyu Hu}
\author[a]{Alina Kiessling}
\author[i]{Nikole Lewis}
\author[d]{Maxime Rizzo}
\author[d]{Aki Roberge}
\author[j]{Tyler Robinson}
\author[k]{Leslie Rogers}
\author[l]{Dmitry Savransky}
\author[h]{Chris Stark}
\affil[a]{Jet Propulsion Laboratory, California Institute of Technology, Pasadena, CA 91109, USA}
\affil[b]{Department of Earth and Planetary Sciences, and Department of Physics, Massachusetts Institute of Technology, Cambridge, MA 02139, USA}	
\affil[c]{University of San Francisco, College of Arts and Sciences, San Francisco, CA 94117, USA}
\affil[d]{NASA Goddard Space Flight Center, Greenbelt, MD 20771, USA}
\affil[e]{Kavli Institute for Particle Astrophysics and Cosmology, Stanford University, Stanford, CA 94305, USA}
\affil[f]{Northrop Grumman Aerospace Systems, Redondo Beach, CA 90278, USA}
\affil[g]{Stanford University, Stanford, CA 94305, USA}
\affil[h]{Space Telescope Science Institute, Baltimore, MD 21218, USA}
\affil[i]{Department of Astronomy and Carl Sagan Institute, Cornell University, Ithaca, NY 14853, USA}
\affil[j]{Department of Astronomy and Planetary Science, Northern Arizona University, Flagstaff, AZ 86011, USA}
\affil[k]{Astronomy Department, University of Chicago, Chicago, IL 60637, USA}
\affil[l]{Sibley School of Mechanical and Aerospace Engineering, Cornell University, Ithaca, NY 14853, USA}

\usepackage{lineno}

\begin{document}

\maketitle

\begin{abstract}
Launching a starshade to rendezvous with the {\it Nancy Grace Roman Space Telescope} would provide the first opportunity to directly image the habitable zones of nearby sunlike stars in the coming decade. A report on the science and feasibility of such a mission was recently submitted to NASA as a probe study concept~\cite{seager19}. The driving objective of the concept is to determine whether Earth-like exoplanets exist in the habitable zones of the nearest sunlike stars and have biosignature gases in their atmospheres. With the sensitivity provided by this telescope, it is possible to measure the brightness of zodiacal dust disks around the nearest sunlike stars and establish how their population compares to our own. In addition, known gas-giant exoplanets can be targeted to measure their atmospheric metallicity and thereby determine if the correlation with planet mass follows the trend observed in the Solar System and hinted at by exoplanet transit spectroscopy data. 
In this paper we provide the details of the calculations used to estimate the sensitivity of {\it Roman} with a starshade and 
describe the publicly available Python-based source code used to make these calculations.
Given the fixed capability of {\it Roman} and the constrained observing windows inherent for the starshade, we calculate the sensitivity of the combined observatory to detect these three types of targets and we present an overall observing strategy that enables us to achieve these objectives.
\end{abstract}




\newpage
\section{Introduction}

Space-based direct imaging is the next frontier of discovery for exoplanet science, moving beyond detection of Earth-like planets toward full characterization of their atmospheres, potentially identifying signs of life.
The {\it Nancy Grace Roman Space Telescope} (hereafter referred to as \wfirst) is scheduled to launch in late 2025.
While its Coronagraph Instrument (CGI) will reach planet-star flux ratios only achievable in space, with the capability to directly image gas giant planets around nearby stars, it does not have the sensitivity to detect Earth-like planets.
With the launch of a companion starshade -- a free-flying 26-m external occulter -- such a goal can be reached within the next decade.

The Starshade Rendevous Probe (SRP) mission concept was presented in a recent report~\cite{seager19} assessing the possibility to fly an external starshade. Using the CGI, the starshade significantly enhances the exoplanet observing capabilities of \wfirst\ 
by providing 1) improved contrast and a reduced inner working angle (IWA), enabling sensitivity to Earth-like exoplanets in the habitable zone,
2) an unlimited outer working angle, enabling a wider view of the planetary system around each target, and
3) higher throughput
(the coronagraph masks are not needed with a starshade),
enabling the sensitivity needed to spectrally characterize an Earth-like exoplanet.

In this paper we provide the technical basis for the SRP study report~\cite{seager19} along with the simulations used to estimate sensitivity of the observatory. While many such assessments have been done in the past~\cite{turnbull12, savransky15, stark14a, stark16}, this study started by defining the science objectives of the mission with detailed knowledge of \wfirst\  constraints, CGI expected performance, and the significant advances of NASA's S5 Starshade technology development program.\footnote{\url{https://exoplanets.nasa.gov/exep/technology/starshade/}} While detailed simulations of Starshade missions exist~\cite{stark16, soto17, soto19}, we found that the SRP has a relatively small number of targets, which imposes different demands. The need to understand how to balance the different objectives of the mission, within significant mission constraints, and how to prioritize targets led us to produce a simulations package focused on the SRP. This software is publicly
available for reproduction of the results below and for comparison with similar simulations.\footnote{\url{https://github.com/afromero/Starshade_Rendezvous_Probe_sims}}

The paper is structured as follows. 
We briefly review the science objectives presented in the SRP report (\S\ref{objectives}) 
followed by and  outline of the top-level mission constraints, based on its partnering with
the \wfirst\ (\S\ref{constraints}).
We then describe the model used to quantify the starshade performance -- both the observatory characteristics and the expected astrophysical scene \S\ref{model}.
Based on this observing plan, we next calculate the expected integration times required to
achieve a given signal-to-noise, given \wfirst's nominal optical design (\S 5).
We use the observing model to calculate the sensitivity of the observatory toward individual targets (\S\ref{sensitivity}) and then provide an overall observing strategy, including choices on the number of stars, the number of visits per star,
and when to switch from imaging to spectroscopy (\S\ref{strategy}).
Last, in \S\ref{results}, we calculate the expected performance of the mission, with a specific list of target stars, summarizing in \S\ref{summary}.

\section{Science Objectives}\label{objectives}

The Starshade Rendezvous Probe Mission Study Report \cite{seager19} developed both the scientific motivation and technical feasibility for such a mission. Further detail on the science case can be found there; here we focus not on motivating the science goals, but rather on whether these goals can be achieved.

In summary, the science objectives of the mission are three-fold:
\begin{itemize}
\item{{\bf Objective 1: Habitability and Biosignature Gases \& The Nearest Solar System Analogs.} {\bf a)} Determine whether Earth-like exoplanets exist in the habitable zones around the nearest sunlike stars and whether have they have signatures of oxygen and water vapor in their atmospheres. {\bf b)} Detect and characterize planets orbiting the nearest sunlike stars.}
\item{{\bf Objective 2: Brightness of Zodiacal Dust Disks.} Establish if the zodiacal cloud of our inner Solar System is representative of the population of our nearest neighbor stars. }
\item{{\bf Objective 3: Gas-Giant Atmospheric Metallicity.} Determine the atmospheric metallicity of known cool giant planets to examine trends with planetary mass and orbital semi-major axis, and to determine if these trends are consistent with our Solar System.}
\end{itemize}
The first objective -- finding Earth-like planets -- is paramount.
As such, maintaining the sensitivity to discover and potentially characterize Earth-like exoplanet candidates around these stars drives the observatory requirements.
Meeting these challenging requirements means the observatory will also be capable of discovering and characterizing a wide range of planet types, 
from those like the giant planets in our Solar System, 
to the sub-Neptune planets commonly discovered by Kepler \cite{borucki11},
and down to Earth-mass planets.

The difficulty of imaging Earth-like planets necessitates a “deep dive” approach;
that is, a detailed investigation of a relatively small sample of our nearest-neighbor sunlike stars --
only the closest sunlike stars where starshade observations have both high imaging sensitivity for exoplanet discovery
and high spectral sensitivity to characterize their atmospheric composition,
while also allowing multiple visits to constrain the orbits of any habitable zone planets found (see \cite{seager19} for a more detailed discussion).
Planetary systems almost certainly exist around several of these stars \cite{winn15}.
If an Earth-like planet candidate is discovered around at least one of these stars,
it will be spectroscopically observed to hunt for water vapor and oxygen.
Three steps are needed to accomplish this goal:
1) initial detection via direct imaging,
2) habitable zone determination, by multi-epoch orbit tracing, and
3) atmosphere characterization, with spectroscopy of the most compelling candidate planets -- particularly those in the habitable zone.
Each of these steps places unique requirements on the mission parameters,
as described in this paper.

The requirement to detect Earth-like exoplanets also enables the ability to detect exozodiacal dust disks and spectral features of gas giants. The distribution of exozodiacal dust brightness, at the level relevant for Earth-like exoplanet detection, is largely unconstrained. Recent bounds on the warm dust disk brightness of many sunlike stars of interest to the SRP are provided by LBTI~\cite{ertel20}, but the sensitivity is still more than an order of magnitude in excess of what is needed. Objective 2 will provide the key information necessary to assess the sensitivity to directly image habitable zone exoplanets. 

Objective 3 will test whether the correlation between atmospheric gas metallicity and planet mass (as well as semi-major axis) observed in our Solar System, and hinted at in a few transit spectroscopy measurements of exoplanets~\cite{wakeford17}, is a universal trend. Identifying this correlation with the SRP would provide evidence of common processes of planetary system formation. 

\section{Mission Constraints}\label{constraints}

The Starshade Rendezvous Probe is an enhancement of the \wfirst\ mission.
As such, many of the starshade mission's parameters are fixed by \wfirst's
established telescope, instruments, and operational timeline.
The key constraints imposed on a starshade mission are summarized in
Table~\ref{missionParamTable}.
\footnote{
Note that all of the parameters in Table~\ref{missionParamTable} are up-to-date at the time of this study. A full list of current \wfirst\ coronagraph characteristics can be found at: \\  \url{https://wfirst.ipac.caltech.edu/sims/Param_db.html\#coronagraph_mode} \label{ipacCGIsite}
}
In particular, \wfirst's Coronagraph Instrument (CGI) will be used
for imaging and spectroscopy.  The bandpass and spectral resolution
for this instrument are designed for a similar science case (exoplanet imaging/spectroscopy)
and, as such, are a good match to our science goals.

The field of regard is limited by both the telescope and the starshade.
\wfirst\ cannot point within 54\degree\ of the Sun or light will scatter
into the telescope assembly.
Solar angles greater than 83\degree\ are excluded by the starshade;
beyond this limit the starshade can no longer be viewed
close to face on without being illuminated by sunlight. Although this last point is not technically an imposed constraint, it is an important one imposed by the Starshade architecture and limits the target observing windows.

The imaging end-to-end efficiency, discussed in more detail in \S\ref{sensitivity}, is the result of losses from the telescope aperture, optical path, with coronagraph masks excluded, and all the way down to the quantum efficiency of the detector. The Starshade observations depend on the CGI camera with little modifications, so we take this as an imposed constraint. The field of view of the CGI detector is limited to 4.5\arcsec\ (radial). This parameter is important for detecting any planets orbiting outside the habitable zone of the target of interest.

Lastly, note that while \wfirst's lifetime requirement is 5 years,
we assume the starshade launch will occur 3 years into the mission,
giving an overlap of 2 years. Extending beyond that duration is ultimately limited by the lifetime of the CGI camera and the overall \wfirst\ system.

\input{missionParams.tex}

\section{Observing Model}\label{model}

\begin{figure}\begin{center}
    \includegraphics[width=6.5in]{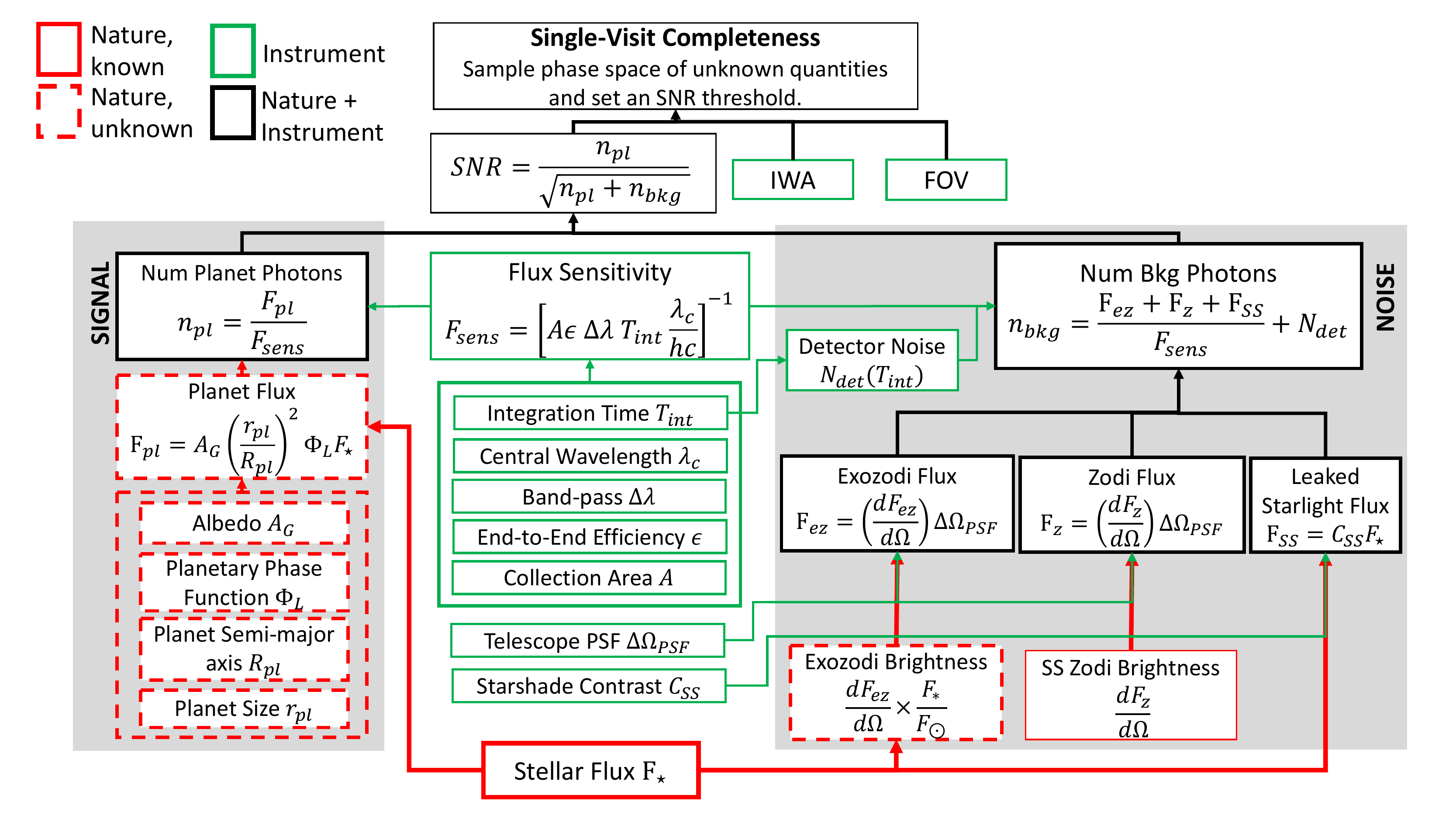}
  \end{center}\caption{
    The detailed observatory model used for the sensitivity estimates in this study. Single-visit completeness is determined by IWA, field of view (FOV), and SNR resulting from a planet observed with the SRP system. Red boxes are parameters entirely determined by nature, green boxes are parameters controlled by observatory design, and black boxes are a combination of both. The same model is used for spectral completeness by using the corresponding values for bandpass and end-to-end efficiency.
  }\label{snrFig}
\end{figure}

This section covers the observing model used to calculate the signal to noise ratio (SNR) of an exoplanet directly imaged with a telescope-starshade system. The overall observing model is shown as a flowchart in Figure~\ref{snrFig}.
This model consists of both astrophysical inputs, shown in the red boxes, (\S\ref{sec:astro_model}) and instrument parameters, shown in the green boxes, (\S\ref{sec:instr_model}), which are combined to estimate the SNR (\S\ref{sec:snr_model}). 

\begin{figure}\begin{center}
    \includegraphics[width=5in]{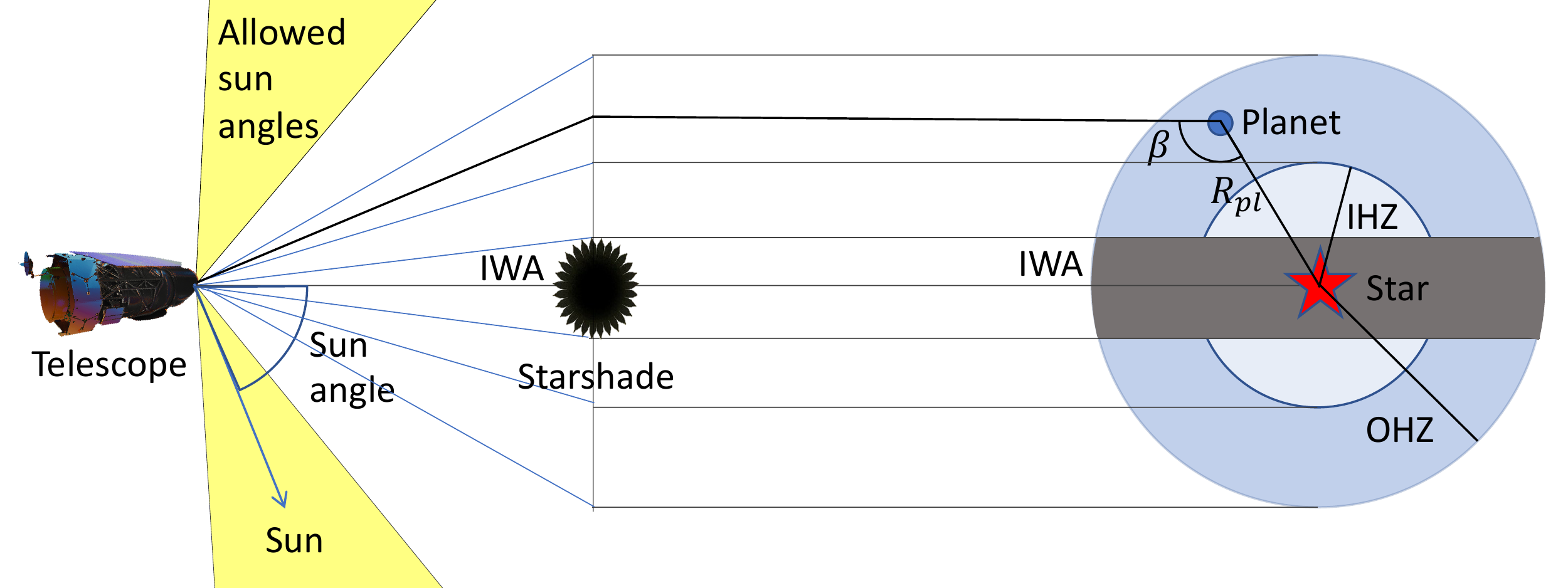}
  \end{center}\caption{
    Starshade observatory geometry. The telescope is pointed at a star with the starshade blocking the starlight. The inner working angle (IWA) is the angle subtended by the direction to the star and the outer radius of the starshade. Observations of planets in this region is excluded. The habitable zone of the star is defined by the inner habitable zone (IHZ) radius and outer habitable zone (OHZ) radius, defined by the distance to the star where the stellar irradiance allows for water in liquid state. A planet at radius $R_{pl}$ has illumination phase angle $\beta$ defined as the star-planet-telescope angle. The telescope-starshade system has a region of Sun angles at which it can observe. The lower limit is set by the solar exclusion angle of the telescope baffle while the upper limit is set by reflection of the Sun off the starshade's surface on the telescope. These solar exclusion angles result in important observational constraints. 
  }\label{obsFig}
\end{figure}

The exoplanet direct imaging observing geometry of the telescope-starshade system is shown in Figure~\ref{obsFig}. The starshade blocks the starlight up to an inner working angle (IWA), above which exoplanets can be observed. A planet at radius $R_{pl}$ is observed with illumination phase angle $\beta$. For habitable zone exoplanets, defined by the inner and outer habitable zone (IHZ, OHZ) radii where water can exist in its liquid phase. The telescope-starshade system has a region of allowed Sun angles over which it can operate with the lower limit defined by the exclusion angle from the baffle of the telescope and the outer limit defined by reflection and scattering of sunlight off the starshade into the telescope baffle. 

\subsection{Astrophysical Model}
\label{sec:astro_model}
The signal is produced by the star's flux density being reflected off the surface of the planet. The astrophysical sources of background light are the exozodiacal dust disk as well as our own Solar system's zodiacal dust. For this subsection, we focus solely on the astrophysical model with the parameters provided in \S\ref{planetParamSection}.
\subsubsection{Star Flux Density}

We use a simple black body model for the star. 
The model inputs are the bolometric luminosity $L_\star$, blackbody temperature $T_\star$, stellar mass $M_\star$, and distance from Earth $d_\star$. The star's spectral radiance is given by Planck's law
\begin{equation}
B_\star(\lambda, T_\star) = \frac{2hc^2}{\lambda^5} \frac{1}{\exp(hc/\lambda kT_\star)-1}.
\label{eq:planck}
\end{equation}
The flux density $F_\star$ at Earth in a band defined by limits $\lambda_\mathrm{min}$ and $\lambda_\mathrm{max}$ is given by
\begin{equation}
F_\star = 
\frac{L_\star}{4\pi d^2_\star}\frac{1}{\sigma_\mathrm{SB} T_\star^4}
\int_{\lambda_\mathrm{min}}^{\lambda_\mathrm{max}} d\lambda \ B(\lambda, T_\star)
\label{Eq:star_flux}
\end{equation}
where $\sigma_\mathrm{SB}$ is the Stefan-Boltzmann constant. The first fractional term in the equation is the 
total wavelength-integrated flux. 
The second fractional term is the normalization by the Stefan-Boltzmann law, which is the integral of Equation~\ref{eq:planck} over all wavelengths.

The input data for the stars used in this study was obtained from ExoCat-v1\footnote{M. Turnbull (2015), ``ExoCat-1: The Nearby Stellar Systems Catalog for Exoplanet Imaging Missions", arXiv:1510.01731,  \url{https://nexsci.caltech.edu/missions/EXEP/EXEPstarlist.html}}, which is a compilation of stars within 30 pc. The target selection is discussed in \S\ref{sensitivity}.

\subsubsection{Planet Flux Density}

For bodies that are not self-luminous, the flux density depends entirely on the light reflected from its surface by the host star with flux density $F_\star$. The fraction of light reflected depends on the ratio of the projected disc area of the planet with radius $r_{\rm pl}$ and square distance from the star $R_{\rm pl}$. The dependence of the reflected light on the phase angle $\beta$ (i.e. the planet-star-observer angle) is assumed to be Lambertian 
\begin{equation}
\Phi_L(\beta) = \frac{\sin\beta + (\pi-\beta)\cos\beta}{\pi}.
\label{Eq:lambertian}
\end{equation}
This assumption of isotropic scattering is approximately correct for cloudy gas giants
\cite{cahoy10, mayorga16}, but not for rocky planets that have enhanced forward scattering.
Our assumed geometric albedo for Earth-like planets (0.2) only matches the Earth's reflectivity for scattering angles $\beta \lapp 90$\degree; for more grazing angles, it is an underestimate\cite{robinson10,robinson18}.
The function $\Phi_L(\beta)$ is normalized to unity at maximum illumination ($\beta=0$), takes the value of $1/\pi$ at maximum elongation ($\beta=\pi/2)$, and vanishes smoothly as the planet eclipses the star ($\beta\to\pi$). The normalization of the reflected light at maximum illumination, taking into account the $(r_{\rm pl}/R_{\rm pl})^2$ dependence and $\Phi_L(\beta)$, is the geometric albedo $A_G$~\cite{robinson11}. Altogether, the flux of a planet is given by
\begin{equation}
 F_{\rm pl}(\beta)
  = A_G \left(\frac{r_{\rm pl}}{ R_{\rm pl}}\right)^2
  \Phi_L(\beta)
  F_{\star}.
\label{Eq:planet_flux}
\end{equation}

The model above applies to the entire range of planet types considered in this study (Earth, Super-Earths, Subneptunes, Neptunes, and Jovian planets) with different parameters. The parameters used for this study are provided in Section~5.

\subsubsection{Dust}

The main sources of natural backgrounds are the sunlight scattered by zodiacal dust within our own Solar System and the starlight scattered by exozodiacal dust surrounding the exoplanet. Our model for the Solar System's (SS) zodiacal dust is from \cite{leinert98}, which considers variations with both wavelength and direction; the ecliptic latitude and longitude of each target star is taken into account. The exozodiacal dust brightness for a 1-zodi disk is set to 22 mag/arcsec$^2$\cite{beichman99a}.  
This is somewhat higher than the nominal brightness of the SS zodiacal dust itself (23 mag/arcsec$^2$) because the exozodiacal dust is twice as thick (we only look through half of the SS zodiacal dust's thickness) and because we view it at more forward-scattering angles \cite{stark14a}.
While some of our target stars do have measured levels of exozodiacal dust, most are non-detections.  Analysis of these upper limits 
suggests a median dust thickness a factor of 4.5 higher than the Solar System
\cite{ertel20}.
We adopt this enhanced level as our fiducial amount of exozodiacal dust.
The exozodi model is scaled based on the ratio of stellar flux to solar flux
and corrected for orbital location relative to Earth-equivalent insolation distance (EEID) given by 1~AU$\times (L_\star/L_\odot)^{1/2}$.

The flux density of SS zodiacal and exozodiacal dust backgrounds are proportional to the solid angle subtended by the point spread function (PSF) core ($\Delta \Omega_{PSF}$), which is inversely proportional to the telescope diameter. For a brightness distribution $dF/d\Omega$, the flux density is approximated by
\begin{equation}
F = \left(\frac{dF}{d\Omega}\right) \Delta\Omega_{PSF}.
\end{equation}

\subsection{Observatory Model}
\label{sec:instr_model}
The observatory is a combination of the \wfirst\ telescope, including the CGI instrument, and starshade occulter. 
\subsubsection{Telescope}

The sensitivity of the telescope can be summarized in a single value $F_{sens}$, which is the flux that would produce a single photon count on average for a given integration time $T_\mathrm{int}$. This is given by 
\begin{equation}
F_{sens} = \left[A\epsilon \Delta\lambda T_\mathrm{int} \frac{\lambda_c}{hc}\right]^{-1}, 
\label{eq:f_sens}
\end{equation}
where $A$ is the geometric aperture of the telescope, based solely on its diameter, and $\epsilon$ is the end-to-end efficiency, which fully accounts for the fraction of photons entering the geometric aperture that produce photon counts in the detector. The product $\epsilon A$ is the effective area of the telescope. The bandwidth is given by $\Delta \lambda$ and $\lambda_c$ is central wavelength, $h$ is  Planck's constant $h$, and $c$ is the speed of light. The detector noise produces a photon-equivalent background count rate given by $N_\mathrm{det}$. The model described in the rest of this section relies heavily on the models used for CGI~\cite{nemati14, nemati17}.

\wfirst\ is a 2.4-m diameter optical telescope, corresponding to a collecting area of $A=4.5$~m$^2$.
The diameter spatial full-width at half maximum (FWHM) resolution is $\theta_{\rm PSF}=0.065$\arcsec  \ at 750 nm. 
The pixel scale for the CGI instrument is 0.0218\arcsec.
The solid angle subtended by the PSF is 
approximated 
as $\Delta\Omega_{PSF}\simeq\pi\theta_{\rm PSF}^2$, which gives $\Delta\Omega_{PSF}\simeq 3.1\times10^{-13}$~sr at 750 nm. 

The Starshade imaging bandpass filter (615-800 nm) is tuned to be sensitive to water vapor and oxygen absorption lines at 720 nm and 760 nm, respectively. 
The imaging field of view of 4.5\arcsec\ (imaging) 
enables observations well beyond the habitable zones of the nearest sunlike stars, providing the potential for discovery of giant outer planets.

The focal plane detector is an electron-multiplying CCD (EMCCD).
The detector noise is $\sim$10 counts/hour, including noise equivalent dark current (dominant term), clock induced charge, and read noise (negligible contribution).
While the EMCCD detector has the advantage of no read noise,
it is significantly degraded by cosmic ray hits;
5 years of degradation is included, reducing the detector quantum efficiency (QE) to 28.5\%
$^{\ref{ipacCGIsite}}$.

The end-to-end efficiency ($\epsilon$), which includes the optical throughput of the telescope and the detector quantum efficiency (dQE), is
3.5\% for imaging and 3.4\% for spectroscopy.
A complete budget for the factors going into these throughput calculations
is provided in Table~\ref{throughputTable}. The \wfirst\ pupil has a central obscuration that results in a reduction of the raw collecting area by 18\%. The light collected at the aperture goes through multiple reflections in various elements of the telescope optics to deliver it to the CGI with a further reduction of 19\%.
Prior to reaching the CGI, a dichroic beam splitter divides the signal into the CGI and a guidance camera with 90\% efficiency. Within the CGI, there are multiple optical elements prior to delivering the light to the detector with an efficiency of 60\%. The detector effective quantum efficiency (QE) is a combination of effects including QE, cosmic rays, and other detector effects, of 28.5\%. 
We use the end-of-life value since the starshade would operate in the last couple of years of the \wfirst\ telescope. The PSF has 34\% of its total light in the core due to the diffraction from the struts and central obscuration region of the \wfirst\ aperture. 
The top-level instrument characteristics are summarized in Table~\ref{missionParamTable}.

While the model in the previous subsection provides an adequate description for the wide-band imaging mode, the spectrograph requires some additional details. The currently planned implementation is a slit-prism spectrograph, which blocks all but a narrow region with a slit of width $D$ and disperses the light orthogonal to the slit's long axis. The key parameter for the design is the slit width $D$ which must be wide enough to accommodate for telescope jitter and motion of the planet during a long period of observation while not being so wide that it allows for a significant amount of zodiacal and exozodiacal dust brightness background photons to disperse into the pixels of interest for the exoplanet. The prism-detector configuration is described by the spectral resolution parameter $R=\lambda/\Delta\lambda$, which can, in general, be wavelength dependent. 
Note that the spectrograph slit limits observations to one planet at a time,
but still allows for simultaneous measurement of the background along the slit.

\input{throughputTable}

\subsubsection{Starshade}

The starshade performance can be summarized by three key parameters --
the inner working angle (IWA),
the instrument contrast ($C_{SS}$), and
the solar exclusion angles
(see Table~\ref{missionPossibilitiesTable}).

\input{missionPossibilities.tex}

The starshade IWA is the angle subtended by the telescope boresight and the outer radius of the starshade.
The starshade design considered here has $\sim$100\% optical throughput at the IWA~\cite{shaklan17}. Although it is, in principle, possible to observe targets at angles below the IWA, the throughput is reduced and the PSF is distorted. For practical purposes, we assume exoplanets are only observable at angles above the IWA.
    The size of the starshade and the distance between the starshade and telescope are determined primarily by the desired inner working angle, the longest wavelength in the observing band, the diameter of the shadow at the telescope, and the required suppression level.  The optical bandwidth, constraints on feature sizes, and factors such as the ratio of the petal length to the overall diameter, and the number of petals, are examples of other factors that also impact the size and separation of the starshade, see \citenum{2007SPIE.6693E..04C, 2011SPIE.8151E..12C, arenberg08, glassman09}).
The IWA is chosen to observe the habitable zones of nearby sunlike stars. An IWA of 100 mas corresponds to a separation of 1 AU at 10 pc, enabling observations of Earth-like planets for solar-type stars within $\sim$10 pc.

The starshade instrument contrast ($C_{SS}$) is defined as the fraction of starlight leaked per resolution element at the IWA. The resulting background flux is $F_{SS}=C_{SS}F_\star$. This results in a background contribution in the habitable zone with $C_{SS}$ specified so as to keep it below the expected zodi and exozodi backgrounds, enabling the sensitivity to detect Earth-like exoplanets. The current best estimate for the contrast is $C_{SS}=4 \times 10^{-11}$ (see~\citenum{seager19} for details). 

In addition to scattered starlight, solar glint and diffracted speckle patterns from the target result in an additional localized, predictable backgrounds that can be calibrated. While solar glint can be a limiting factor, recent progress in petal edge design indicates that it can be suppressed outside the IWA to negligible levels. We therefore do not include a treatment of solar glint in this study; for a more detailed treatment see \citenum{hu20}. The starshade speckle pattern and its ability to remove it during analysis are not included in this study. It is expected that they become an important source of backgrounds These effects will be accounted for in more detail with more recent tools such as the SISTER simulations package and upcoming Starshade data challenges. 

The starshade can accommodate a relatively wide bandpass (26\%), compatible with the bandpass of the CGI (Table~\ref{missionParamTable}). The relation between starshade design and bandpass is discussed in~\citenum{2011SPIE.8151E..12C}.

One important observing constraint is that, with a starshade, the telescope pointing is limited to at most 83\degree\ from the Sun. Beyond this solar exclusion angle, the starshade 
reflects a significant amount of sunlight into the telescope.

\subsection{Signal-to-Noise Ratio Model}
\label{sec:snr_model}

The signal to noise ratio $SNR$ is approximated as a function of the number of photon counts from the planet of interest $n_{\rm pl}$ and the sum total contribution of background photon counts $n_{\rm bkg}$ according to 
\begin{equation}
SNR = \frac{n_{\rm pl}}{\sqrt{n_{\rm pl}+n_{\rm bkg}}}
\end{equation}

The conversion of the planet flux $F_{\rm pl}$ to photon counts $n_{\rm pl}$ is given by the observatory's single photon flux sensitivity $F_{\rm sens}$ (Equation~\ref{eq:f_sens}) according to $n_{\rm pl}=F_{\rm pl}/F_{\rm sens}$. 

The background photon counts, $n_{\rm bkg}$, have two main contributions, the external background fluxes (leaked starlight, zodiacal and exozodiacal light) and the detector noise counts $N_{\rm det}$. The external background fluxes are converted to background photon counts via the relation to the observatory's single photon flux sensitivity to give
\begin{equation}
n_{\rm bkg}=\frac{F_{ez}+F_z+F_{SS}}{F_{\rm sens}}+N_{\rm det}
\end{equation}

The modifications to the SNR estimate for spectroscopy are listed as follows. 
The bandwidth now corresponds to each sub-band of the spectrometer, which is given by $\Delta\lambda=\lambda/R$. With a characteristic value of $R\sim50$, the bandwidth is at the central wavelength of 750~nm is 15~nm instead of 185~nm. This reduces the single photon flux sensitivity $F_{sens}$ by roughly an order of magnitude compared to imaging mode. 

The leaked starlight as well as the zodiacal and exozodiacal emission 
can be dispersed into the planet spectrum.
If $\theta_\mathrm{slit}$ is the angular width of the slit, the effective increase in the background photon counts $n_{bkg}$ is a factor of $\theta_\mathrm{slit}/\theta_{PSF}$ 
\begin{equation}
n_{bkg}=\frac{\theta_{slit}}{\theta_{PSF}}\frac{F_{ez}+F_z+F_{SS}}{F_{sens}}+N_{det}
\end{equation}
under the assumption that $\theta_\mathrm{slit}\geq\theta_{PSF}$. A slit width of 120 mas is assumed for the Starshade slit prism spectrometer. The reason the slit has to be wider than the PSF core of 65~mas is to accommodate the a priori unknown motion of an Earth-like exoplanet over the data lag of several days nominal spectral integration time period of 25 days. We are assuming that during a spectroscopic observation the slit position can be adjusted over a period of several days given data telemetry, analysis, and commanding latencies. 

The estimates made here assume that the leaking starlight, stray light from the starshade, and exozodi light have been approximated using smooth distributions. The treatment of deviations from these assumptions requires more sophisticated imaging simulation tools (such as SISTER) and exploring a wider range of second order corrections to be considered. These will be treated in future mission concept studies. 

\section{Target Sensitivity}\label{sensitivity}

The analysis presented in this section estimates the performance of the observatory on a per-target basis and has not assumed any constraints on retargeting time or total mission duration, which are covered in the next section. This serves as an initial bound on how many targets the observatory is sensitive to and sets a clear goal for the more complicated problem of visit strategy and retargeting maneuvers. 
We divide the discussion between three types of targets,
corresponding to the three objectives described in \S\ref{objectives}:
1) potential Earth-like planets orbiting bright nearby stars (\S 5.1), 
2) known exoplanets from radial velocity measurements at wide angular separation from their host star (\S 5.2),
 and
3) dust disks that may surround any of the target stars (\S 5.3).

\subsection{Sensitivity to Earth-like Planets}\label{earthTargets}

\subsubsection{Parameters}\label{planetParamSection}

The parameters that define the Star's flux density (Equation~\ref{Eq:star_flux}) and mass are taken from ExoCat \cite{turnbull15}. Table~\ref{targetList-HZ} shows the star parameters for targets selected by the procedure defined later in this section.
The terrestrial planet parameters in our model (Equation~\ref{Eq:planet_flux}) are assumed to have an Earth-like geometric albedo of $A_G=0.2$, based on distant observations of Earth with the EPOXI spacecraft~\cite{robinson11}, along with the Lambertian (isotropic) scattering phase function in Equation~\ref{Eq:lambertian}.  The range of planet radii considered is bounded above at $r_{\rm pl}\leq 1.4$ $\REarth$, based on evidence that suggests that planets with radius below  are predominantly rocky \cite{rogers15}. The lower bound on terrestrial planet radii depends on the planet's ability to retain an appreciable atmosphere, which, in turn depends on their stellar illumination. This results in a dependence on the planet's semi-major axis $R_{\rm pl}$, modified by the stellar luminosity to give $r_{\rm pl}/r_\oplus \geq 0.8 (R_{\rm pl}/{\rm AU})^{1/2}  (L_\star/ L_\odot)^{-1/4} $ \cite{gaudi20}. These maximum and minimum radii serve as the defining limits for Earth-like planets in the simulation results presented here.
The adopted ranges of planetary parameters for habitable zone planets 
(defined here) and gas giant planets (\S\ref{sec:gas_giants} below) are summarized in Table~\ref{planetParamTable}.

The orbital location of Earth-like planets is defined as the habitable zone (HZ) -- the region around a star where a rocky planet with a thin atmosphere may have liquid water on its surface. The location of the habitable zone depends both on the stellar luminosity and on assumptions for the planet's cloud properties.
We adopt a conservative estimate for the habitable zone orbital radii $R_{pl}$ from 0.95 to 1.67 AU for a Solar-luminosity star \cite{kasting93,gaudi20}.
These orbital radii scale by the square root of the
stellar luminosity, to keep the same insolation range as the Solar System.

\input{planetParams}

Planet sizes and semi-major axes are drawn randomly from these defined ranges
for Earth-like planets, based on the distribution defined by SAG-13~\cite{belikov17} and modified by HabEx to include the dependence of the orbital semi-major axis on the lower limit of planet radii. This is determined by drawing from the distribution defined by
\begin{equation}
    \frac
    {\partial^2N(r_{\rm pl}, P)}
    {\partial\ln r_{\rm pl} \ \partial\ln P}=0.38 \, r_{\rm pl}^{-0.19} \, P^{0.26}
    \label{eq:planet_distrib}
\end{equation}
where the orbital period $P$ defines the orbital radius $R_{\rm pl}$ by way of the stellar mass $M_\star$ using Kepler's third law.
For Earth-like exoplanets, the orbits are assumed to be circular,
consistent with most previous studies, e.g.\ \citenum{stark16}.
The estimates made on target sensitivity take into account target availability windows (\S\ref{habitability_windows}) determined by solar exclusions angles along with the Keplerian motion of the planet and the associated changes in planet brightness.
 
For the exozodiacal dust environment, we assume a constant fiducial value of 4.5 zodi based on median dust thickness a factor derived from LBTI limits and measurements~\cite{ertel20}. Two targets of interest, epsilon Eridani and Vega, have measurements of warm dust disk brightness of 300 zodi and 33 zodi, respectively. 
These values are well in excess of 10~zodi, which
significantly increases the integration time for detection of Earth-like planets
and the risk of contamination from planet-induced disk structure \cite{stark08,defrere12}.
As such, these targets have been removed from the list that was presented in the Starshade Rendezvous Probe study report~\cite{seager19}. 

\subsubsection{Treatment of Binaries}\label{binarySection}
\begin{figure}[ht]\begin{center}
    \includegraphics[width=4in]{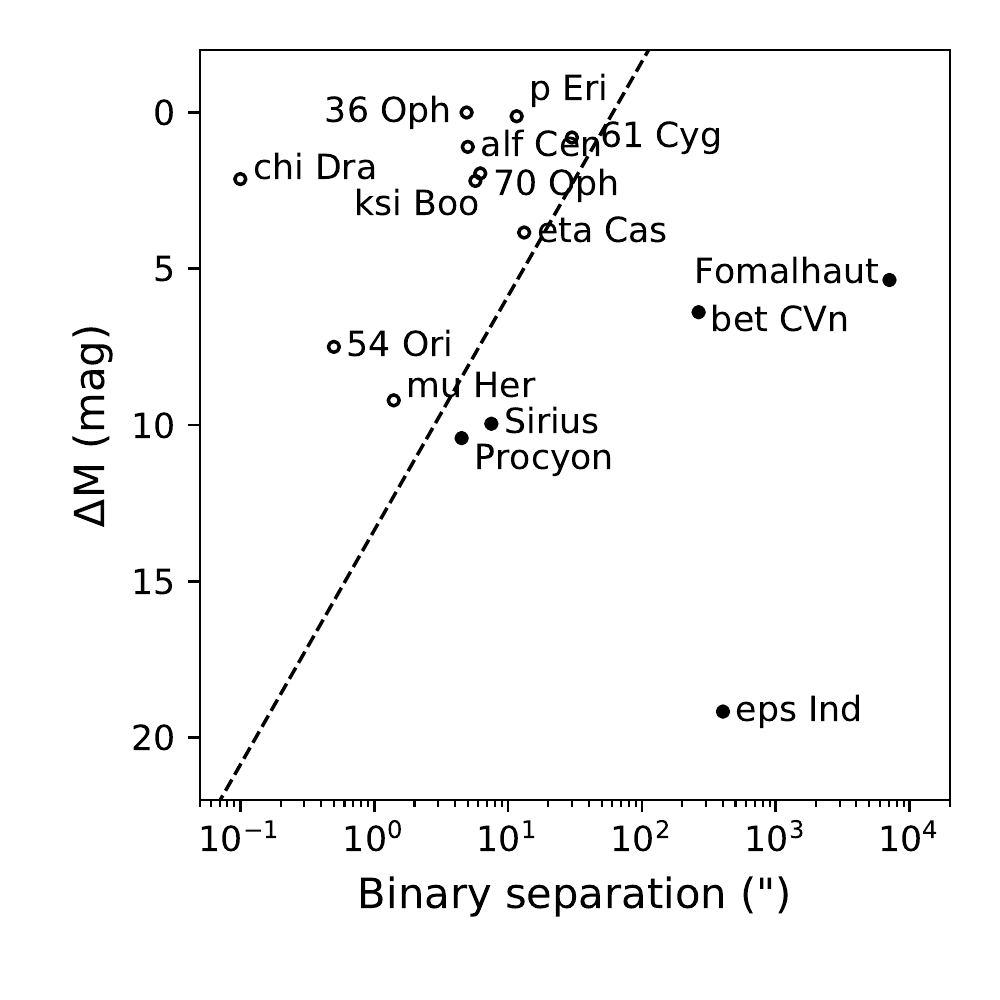}
  \end{center}\caption{
    Angular separation and difference in brightness are shown for all binaries
    that are potential targets.
    Diffraction from the secondary star produces a background contrast level of 
    $4 \times 10^{-11}$ (i.e.\ comparable to our instrumental performance) 
    along the dashed line.
    Binaries with high levels of binary contamination (i.e.\ those to the left of the line) are shown as open circles,
    while those that are still viable targets are filled circles.
  }\label{binaryTargets}
\end{figure}

Nearby optical companions to potential target stars can create
light leakage comparable to the starshade instrument contrast depending on their relative brightness and separation.
Diffracted flux at an angle $\Theta$ away from a companion star can be approximated as 
$F/F_0 \simeq 4 / (\pi x^3)$, where $x \equiv \pi \Theta / \lambda D$.
(This formula is just the large-angle approximation for an Airy diffraction pattern; the error in this approximation is $<$1\% beyond the third Airy ring.)
Figure~\ref{binaryTargets} shows the angular separation and
difference in magnitude for nearby binary stars
(those with $V <$ 5 mag and distance $<$ 8.5 pc).
Diffraction from the secondary star creates additional background flux near the primary
(shown for $10^{-11}$, $10^{-10}$, and $10^{-9}$ contrast levels).
Stars with excessive levels of background (relative to the starshade contrast
floor of $4\times 10^{-11}$) are dropped from the target list (open circles in Figure~\ref{binaryTargets}). We note that mu Hercules was in the target list used for the Starshade Rendezvous Probe study report~\cite{seager19} but is removed from this updated list due to contamination from its nearby optical companion.
Some wide binaries still remain as viable targets
(e.g.\ Procyon, with a 10-magnitude-fainter white dwarf companion at
4.3\arcsec\ separation), shown as filled circles.
While these companions are typically at 100's of AU separation,
Procyon B orbits at only 15 AU (with periapse of 9 AU); this relatively
close orbit could impact the formation and evolution of habitable zone planets.

Note that the background contamination considered here is idealized as 
solely due to diffraction.  Optical aberrations will contribute additional scattering.  For the \wfirst\ telescope, \citenum{sirbu17} estimate that these aberrations could increase the effective contrast limit by $\sim$1--2 orders of magnitude, 
such that borderline systems (Procyon and Sirius) would have their imaging performance significantly degraded.
Very wide/faint binaries (Fomalhaut, eps Ind, bet CVn) still have insignificant contribution, even with the telescope aberrations included.

\subsubsection{Background Contributions}
\begin{figure}\begin{center}
    \includegraphics[width=5in]{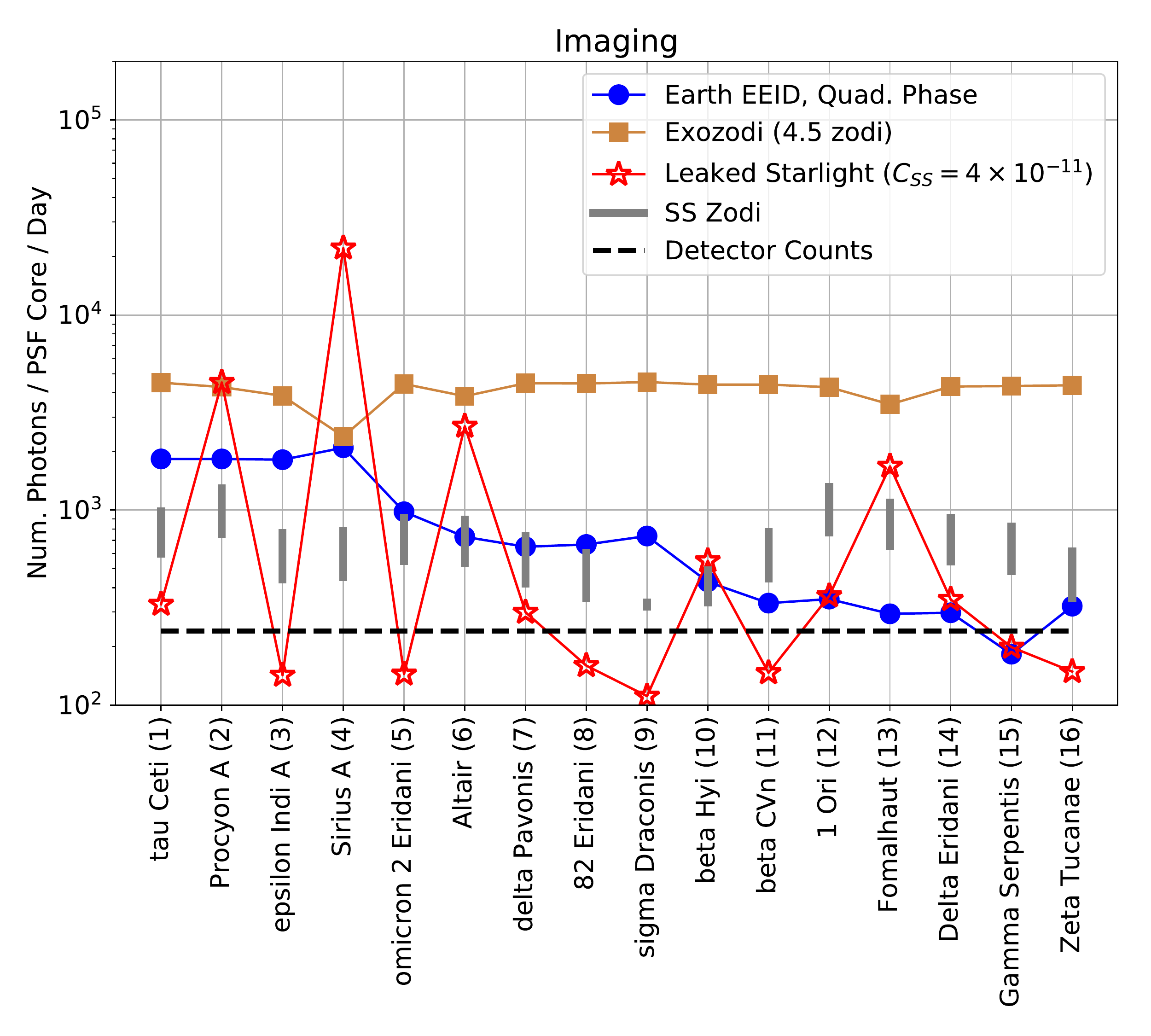}
  \end{center}\caption{
    Photon counts due to each contribution in the SNR model for the 16 nearby stars that provide the best sensitivity to Earth-like exoplanets in their habitable zones. While the Earth-like planet flux densities can vary widely depending on radius, distance to the star, and illumination phase angle, we provide estimates for a planet with Earth parameters at EEID in quadrature phase. In most cases, the exozodiacal dust disk brightness, with a fiducial brightness of 4.5 zodi, dominates the background photon contribution. The leaked starlight flux, assuming an instrument contrast $C_{SS}=4 \times 10^{-11}$ is, in most cases, comparable to the Solar System zodiacal (SS Zodi) dust brightness. The vertical bars of SS Zodi represent the variation depending on the time of year the target is observed. The detector noise is not a major contributor to the error budget. 
  }\label{Fig:photon_counts}
\end{figure}

Each target has several sources of background noise that limits the sensitivity to Earth-like exoplanets. It is illustrative to show the relative contributions from each source as this determines which targets are viable for habitable exoplanet observations. 
In Figure~\ref{Fig:photon_counts} we show the contributions of photon counts, assuming 1 day of integration time in imaging mode under the following assumptions. We assume an Earth-like exoplanet at EEID in quadrature phase. The exozodiacal dust disk brightness has a fiducial value of 4.5~zodi. The leaked starlight assumes an instrument contrast $C_{SS}=4\times10^{-11}$ everywhere, which is conservative since the leaked starlight generally decreases away from the IWA. 
The Solar System zodi background is shown as bars indicating the range of values it can take depending on when the observation is made. Finally, the detector noise contribution is shown as a dashed line. 
The targets are ordered by the brightness of the Earth-like exoplanet. For most targets, the exozodiacal dust disk brightness dominates the background followed by the Solar System's zodiacal dust disk brightness. The leaked starlight at the inner working angle can be stronger in cases where the star is very bright. Note, however, that for these stars the habitable zone will be pushed out to radii typically much higher than the IWA, where the leaked starlight drops. The detector noise counts lie below the contribution of Solar System's zodiacal dust. 

The target selection is based on search completeness, discussed in more detail the next subsection, which depends on the target availability windows, the field of view available around the star, and the range of parameters sampled for terrestrial exoplanets. While the photon counts estimated in Figure~\ref{Fig:photon_counts} do not capture all these details, they do provide a sense of which targets will provide the highest sensitivity to Earth-like exoplanets.

\subsection{Observing Windows}\label{sec:windows}

\begin{figure}[ht]
\begin{center}
    \includegraphics[width=5in]{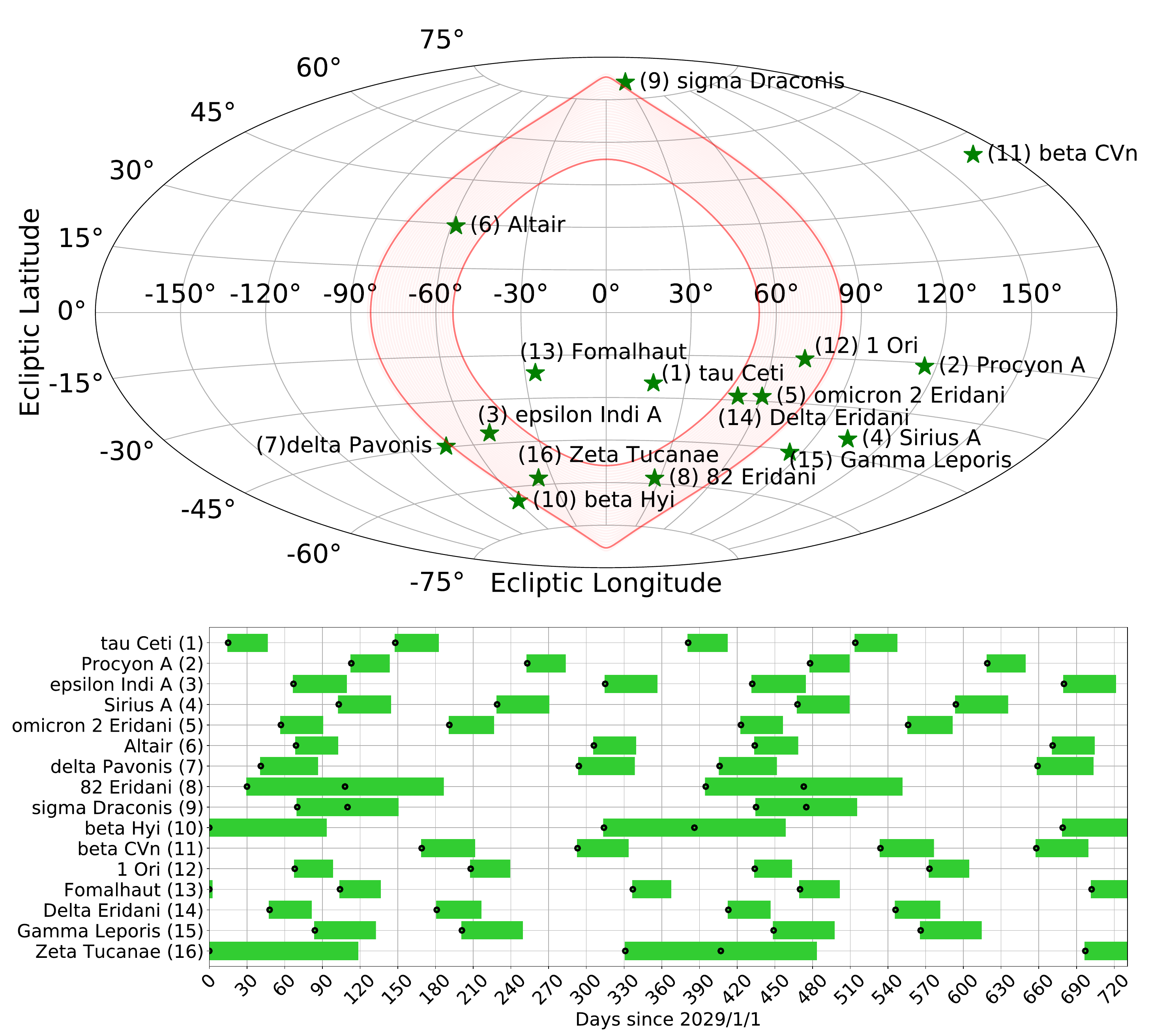}
  \end{center}\caption{
    Top: Sky positions of the Habitability and Biosignature Gases targets in ecliptic coordinates. An instantaneous observing region due to solar exclusion angles (54\degree\ and 83\degree, respectively) is shown as a light red shaded region centered on $0^\circ$ ecliptic longitude.
    Bottom: Target star observing windows as constrained by telescope and starshade solar exclusion angles. These windows result from the instantaneous observing region in the panel above shifting in ecliptic longitude with a yearly period. Each star typically has two $\sim$30-day-long observing windows per year, while higher-latitude stars have a single observing window per year that is longer in duration.  For the sample of nearby stars to be searched for Earth-like planets (upper panel), the black dots correspond to the desired observation start times, to allow for sufficient time for a spectral characterization. 
  }\label{habitability_windows}
\end{figure}

The target availability windows with the Starshade/\wfirst \ system is an important constraint on the observatory's ability to spectrally characterize and determine the orbits of Earth-like exoplanets. 
The Starshade/\wfirst\ system can only observe stars between 54\degree\ and 83\degree\ from the Sun.
For a star in the ecliptic plane, this limited visibility results in two observing windows per year, each $\sim$30 days long.
With increasing ecliptic latitudes the windows become significantly longer, until they merge at $54^\circ$ to produce a single yearly window lasting several months long, then decreasing until it vanishes above $83^\circ$.
Stars very close to an ecliptic pole (e.g.\ chi Dra at 83.6\degree\ ecliptic latitude) are never observable.
The sky position and observing windows for all of our habitability and biosignature targets are summarized in Figure~\ref{habitability_windows}, divided into the nearby-star planet search (upper panel) and the known-exoplanet sample treated in \S 5.2 (lower panel). The sky position and observing windows for all of our gas-giant atmospheric metallicity targets (treated in \S\ref{sec:gas_giants})
 are shown in Figure~\ref{habitability_windows}.

\begin{figure}[ht]
\begin{center}
    \includegraphics[width=5in]{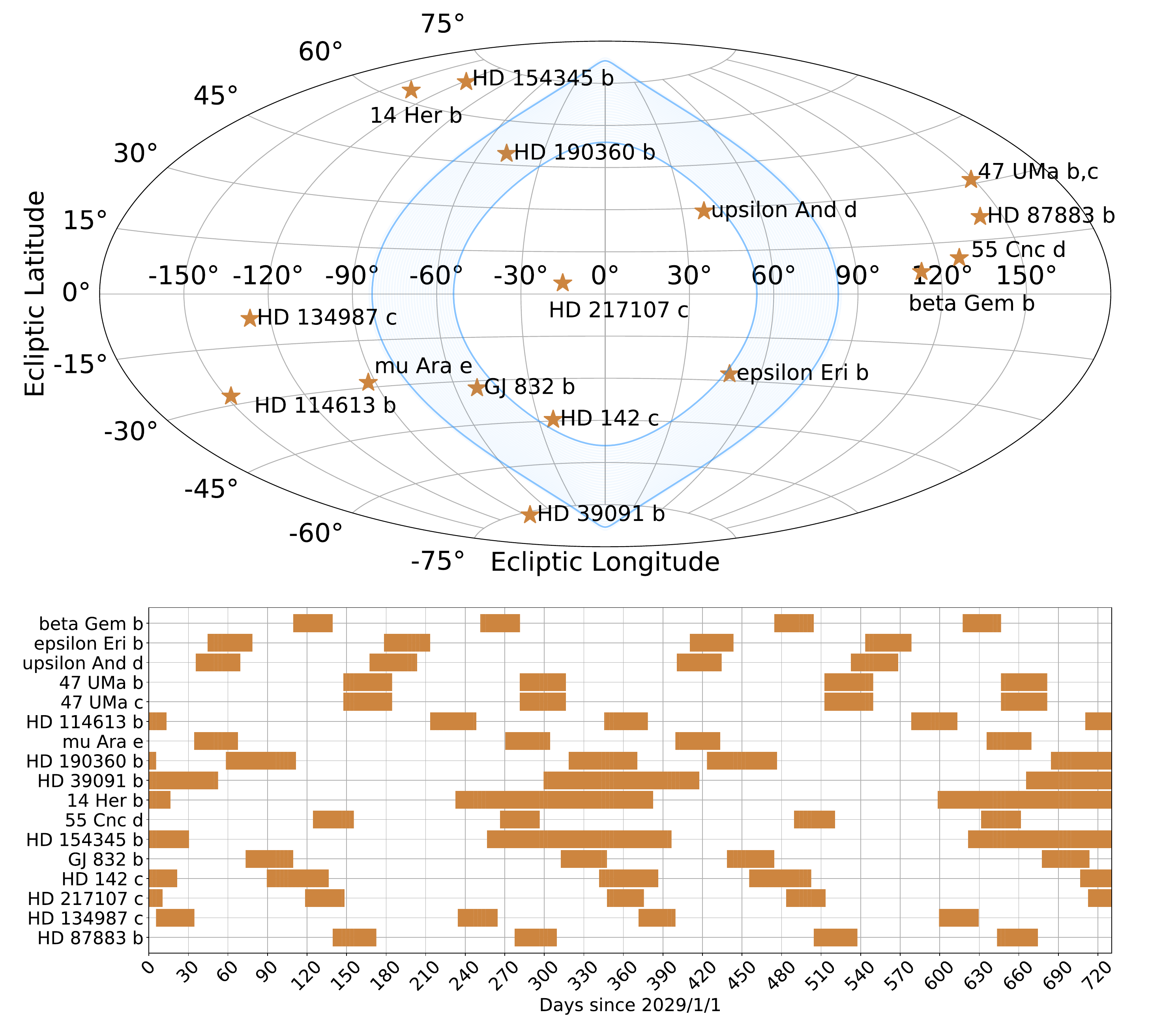}
  \end{center}\caption{
    Top: sky position of known exoplanets for the gas-giant atmospheric metallicity investigation in ecliptic coordinates (\S\ref{sec:gas_giants}). An instantaneous observing window due to solar exclusion angles (54\degree\ and 83\degree, respectively) is shown as a light blue shaded region centered on $0^\circ$ ecliptic longitude. Bottom: Target star observability windows as constrained by telescope and starshade solar exclusion angles.  
  }\label{metallicity_windows}
\end{figure}

The limited observing windows for each target
provide a primary constraint on our observing strategy.
During the 2-year lifetime of the mission, there will
generally be 4 opportunities to observe each target.
While spectral characterization can be performed with only a single visit with favorable illumination phase, 
multiple epochs are needed to constrain the planet's orbit,
in particular its semi-major axis.
The semi-major axis indicates the average amount of stellar radiation
received from the parent star
and thereby determines whether the planet is in the habitable zone.

\begin{figure}[ht]\begin{center}
    \includegraphics[width=6in]{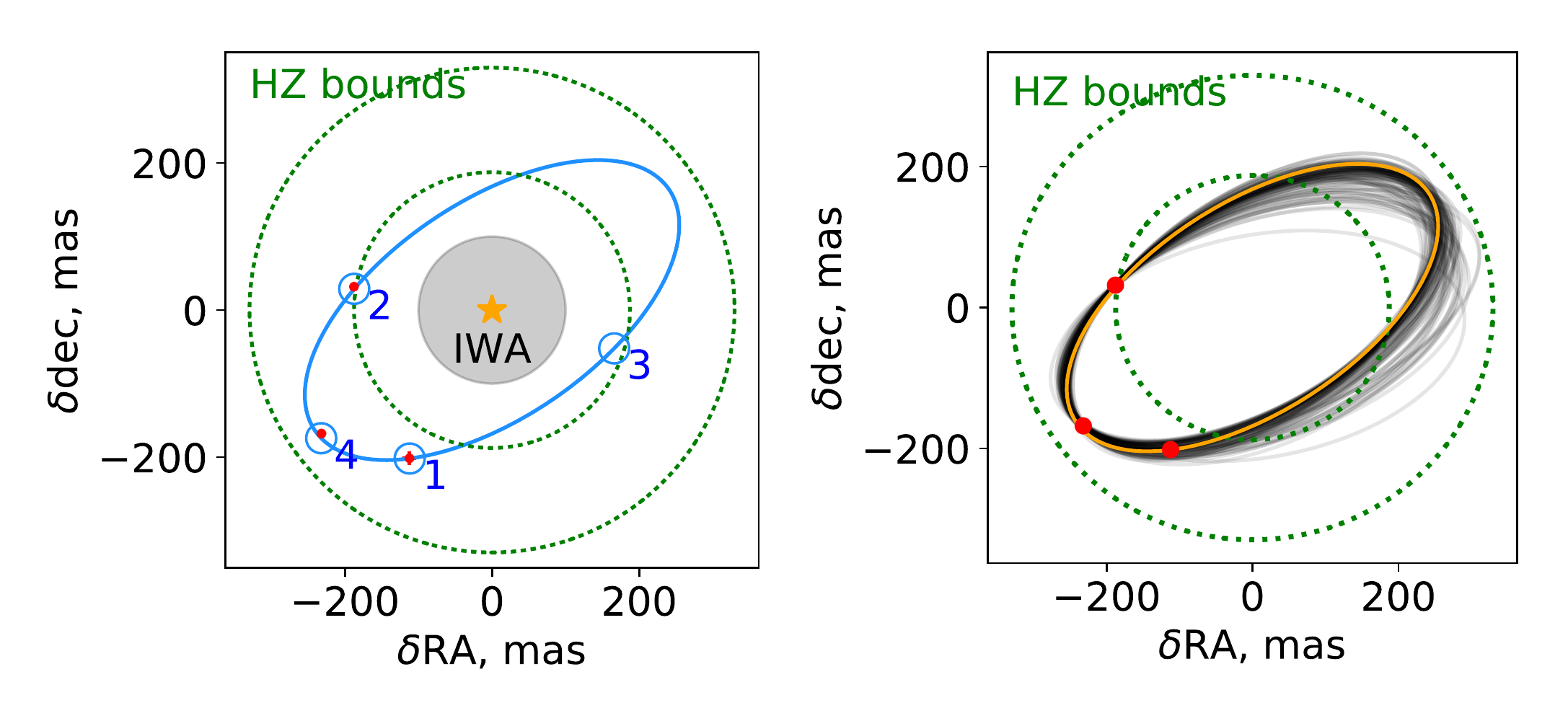}
  \end{center}\caption{
 Four visits of Tau Ceti, as planned with Starshade Rendezvous Probe (Figure~\ref{habitability_windows}), with three detections are sufficient to constrain orbits to the habitable zone (shown in dashed green lines). The left panel shows a simulation of an Earth-like planet with a circular orbit. True positions marked in blue circles and astrometric estimates with error bars shown in red. The numbers indicate the visit number for each observation. 
 The inner working angle (IWA) is shown in gray.
 On the right, sample orbits (gray lines) from a MCMC-reconstructed posterior distribution demonstrate that the fit is well within the HZ, including uncertainties in the orbit eccentricity. The true orbit is shown in orange. 
  }\label{orbitFit}
\end{figure}

Determining the semi-major axis with sufficient accuracy requires at least three astrometric measurements of the planet’s position spread out over two years.
An example of an orbit reconstruction simulation is shown in Figure~\ref{orbitFit}.
Earth-like exoplanets in the HZ are generated with randomly sampled Keplerian orbital parameters with the planet’s phase-varying brightness and associated astrometric precision, defined as the telescope resolution divided by imaging SNR.
The simulated observations are then reconstructed with a Markov chain Monte Carlo (MCMC) that forward models the simulated data.
An ensemble of these simulations for each of our target stars
demonstrates that Earth-like planets can typically be constrained
to the habitable zone with $>$80\% confidence \cite{paper2}.

\subsubsection{Search Completeness and Target Selection}
Our primary metric for evaluating the observatory performance for each star is target completeness -- the fraction of habitable zone planets that can be completely characterized. In order to determine whether a planet is habitable, we need to be able to detect it, constrain its orbit to know that it is indeed in the habitable zone, and take a spectral measurement to determine whether it has an atmosphere with biosignature gases. 
We therefore estimate the following completeness values:
\begin{itemize}
{\item 1) {single-visit completeness}, the fraction of habitable zone planets
that can be effectively imaged at any one time
(defined as SNR$>$7 within a 1-day integration)\footnote{ Note that the SNR threshold was changed from 5 in the Probe Study Report~\cite{seager19} to 7 in this study to reduce the probability of false positives. This values is consistent with the detection threshold used in the HabEx study report.},}
{\item 2) orbit determination completeness, the fraction of observed planets whose orbits
are in the habitable zone (assuming 4 observing epochs),}
{\item 3) spectral characterization completeness, fraction of imaged planets whose
 spectra can identify key atmospheric constituent (SNR$>$20 within a 25-day integration), and}
{\item 4) target completeness, the fraction of observed planets that meet conditions 2 and 3 above. }
\end{itemize}

The single visit completeness serves as a first cut to identify targets where Earth-like exoplanets have a high probability of being detected\cite{brown05}. It is important to note that if a planet is not detected in a single visit, it does not mean it is absent since single visit completeness with a Starshade and CGI is $\lesssim 0.70$~(Table~\ref{targetList-HZ}). It is equally important to note that a single detection of a planet in a region consistent with the habitable is not enough to conclude that it is indeed a habitable zone exoplanet. Follow-up observations that constrain the planet's orbits are necessary to determine that.

The orbit determination completeness is the probability that the planet's orbit can be constrained to be in the habitable zone. In a separate study \cite{paper2}, 
it was determined that 3 detections in 4 visits to the target was sufficient to constrain the orbit of a habitable zone exoplanet with $>$80\% confidence,
depending on the orbital inclination and the phase of observation.
The orbit determination completeness, in this study, is the probability that at least 3 detections occur with 4 visits. 
The simulations in that study sample the orbit periods and observation windows assumed here and performs a Markov Chain Monte Carlo fit of the observations to estimate the posterior distribution of the planet's semi-major axis. 
The number of visits is limited by the lengths and periodicity of the target availability windows for most stars of interest (see Figure~\ref{habitability_windows}).

The spectral characterization completeness is the probability that a spectroscopic observation of a target is successful in any one of 4 visits. The criteria of success (SNR$>$20) is based on a study by \citenum{feng18}, which found that this is was the minimum needed for detection of molecular oxygen and water vapor lines in the CGI band. The 25-day integration time window is the typical maximum for most targets, although some are available for significantly longer.

\begin{figure}\begin{center}
    \includegraphics[width=5in]{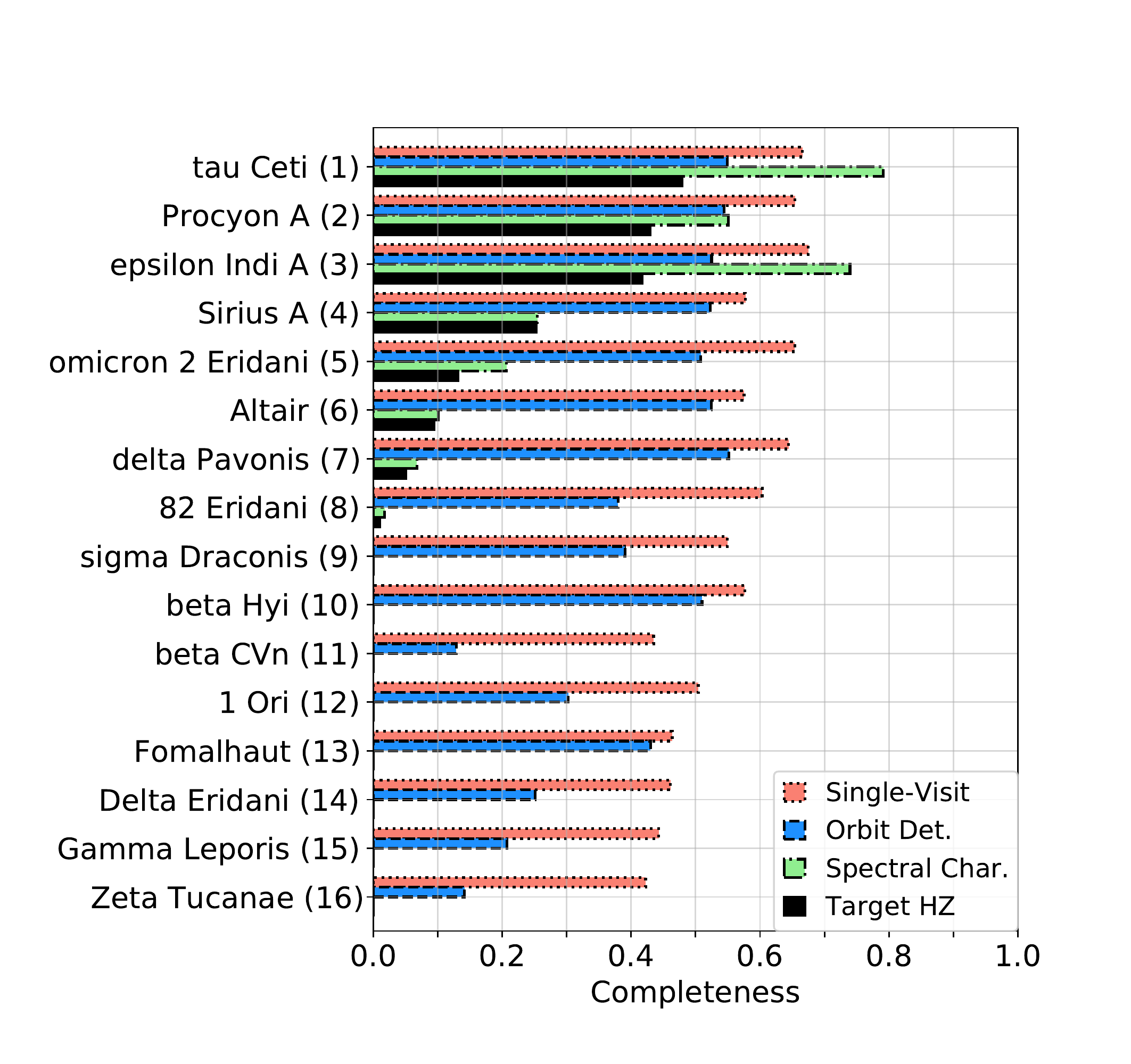}
  \end{center}\caption{
  Completeness estimates are shown for the detection of Earth-like planets around nearby stars.
The simulation results are sorted from best target completeness (top) to least (bottom).
The overall target completeness is shown in black, while its composite factors
are the single-visit completeness (orange),
orbit determination completeness (blue),
and spectral characterization completeness (green). 
  }\label{completeness}
\end{figure}

The target completeness requires all criteria above are met; it is the probability that the orbit constraint requirements and a spectral observation is achieved for a nearby star. 
For each system, we calculate the completeness with a Monte Carlo sampling of 
habitable zone orbits. 
We sample random semi-major axes (using Equation~\ref{eq:planet_distrib}), 
orbital inclinations (cosine distributed), and true anomaly (uniformly distributed for a circular orbit). 
Circular orbits are assumed. Each randomly selected planet is propagated along its Keplerian orbit, with the time of observation limited to 4 observing windows spaced over 2 years (see Figure~\ref{habitability_windows} in \S\ref{sec:windows}). 
A list of 16 stars for finding Earth-like planets are listed in Table~\ref{targetList-HZ} and summarized in Figure~\ref{completeness}, which shows simulation results
for single visit, orbit determination, spectral characterization, and the overall target completeness for each targets. 

\input{targetList}

Most of the habitable zone is visible for all the targets, with single-visit completeness
ranging from $\sim$0.5 to $\sim$0.7 (red bars in Figure~\ref{completeness}). 
Most of the planets that are visible in their habitable zone can also have their orbits traced over multiple epochs, resulting in orbit determination completenesses of $\gapp$0.5 for most stars (blue in Figure~\ref{completeness}). The ability of the observations to constrain each planet's orbit will be described in a companion paper \cite{paper2}. 

The predominant limiting factor is the spectral characterization completeness (green in Figure~\ref{completeness}), which varies by orders of magnitude between stars. Only the brightest stars provide enough photons to produce a high-quality reflected light spectrum within the integration time limits.
For the best targets, spectral characterization completeness can be as high as $\sim$0.8 but as the expected planetary reflected light flux density decreases, 25 days is not sufficient to achieve a spectral SNR$>$20.
Although these fainter planets will not meet our primary science objective, the probability of detection and orbit constraint is still very high, and lower-SNR spectra will still be sensitive to some atmosphere types. This will be the subject of a future investigation.

Of the total of 8 stars that have non-zero target completeness, tau Ceti has the largest ($\gapp0.5$). The remaining targets still have a significant completeness for detection and orbit determination of Earth-like exoplanets. 
These stars are of interest for reconnaissance of planets in orbit and for observing their exozodiacal dust disk brightness in preparation for more sensitive observatories in the future.

\input{targetList_RVplanets_tauCeti}
Note that tau Ceti has two already-discovered super-Earth planets 
that are widely separated and bright enough to have their atmospheres characterized \cite{feng17}.  
Integration times to obtain spectra are listed in 
Table~\ref{targetList-tauCeti} as a function of each planet's illumination phase.

Other than the observing window limitations on integration time, the analysis presented here is on a per-target basis and has not assumed any constraints on retargeting time or total mission duration.
Nevertheless, there is a effective upper bound on number targets
set by our requirements (orbit determination and spectral characterization).
While the number of target stars could be increased with a greater allocation of telescope time, the integration time needed to achieve a successful spectral measurement is limited by the solar exclusion angles.
The Roman Space Telescope CGI with starshade is therefore more limited by sensitivity than it is by telescope time allocation. 

\subsection{Sensitivity to Gas-Giant Planets}
\label{sec:gas_giants}

\begin{figure}\begin{center}
    \includegraphics[width=4in]{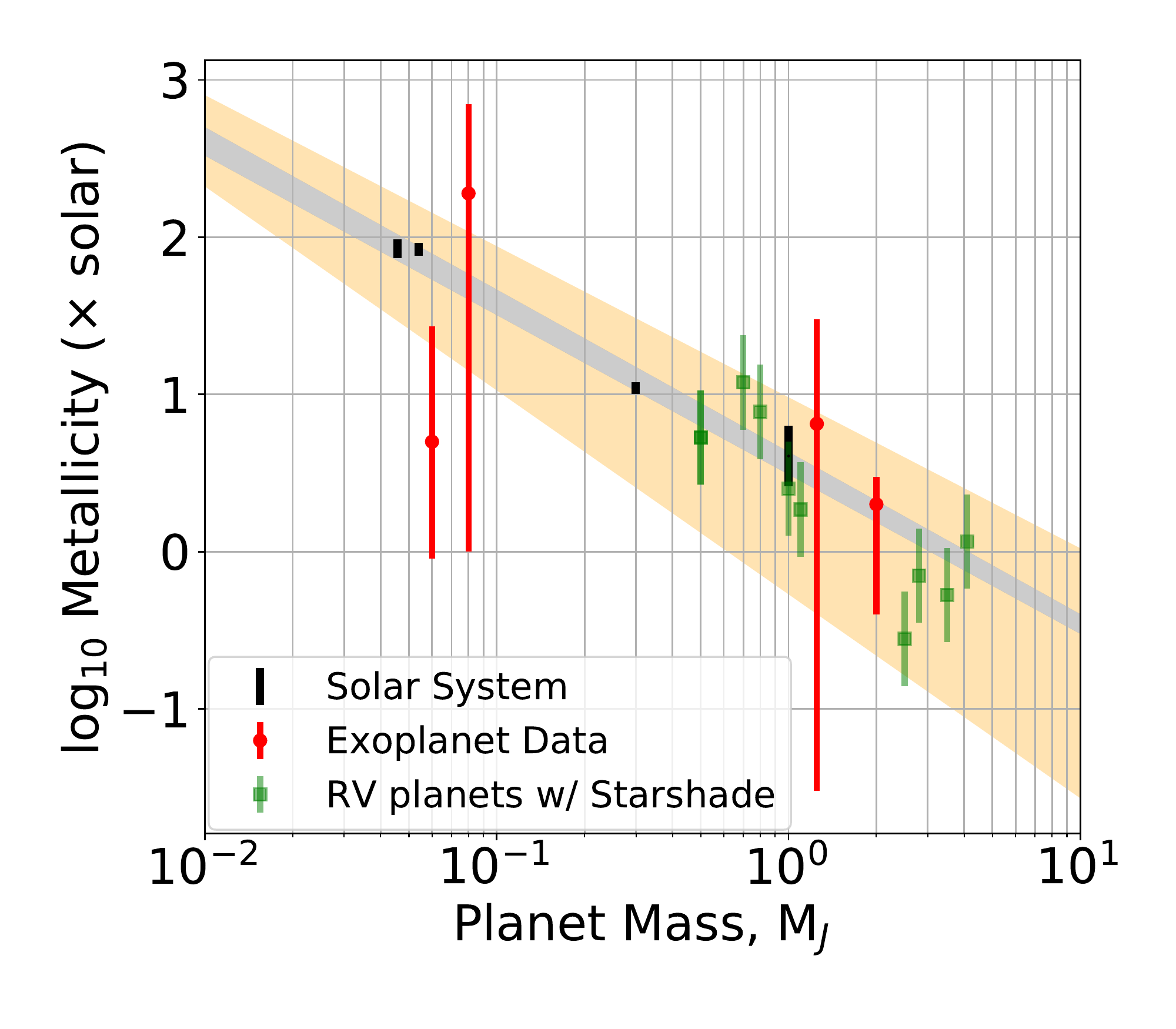}
  \end{center}\caption{
    The correlation of atmospheric metallicity and planet mass. The data, with uncertainties, is shown for the Solar System (black bars) and exoplanet transit spectroscopy measurements~\cite{wakeford17} (red bars). 
    The green points show a random sample of 10 known gas giant exoplanets that could be observed with Starshade Rendezvous Probe assuming they follow the same correlation with 30\% fractional uncertainty in atmospheric metallicity. 
  }\label{metallicity_mass_correlation}
\end{figure}

Next we consider known gas-giant planets as targets for atmospheric characterization.
The goal here is to determine whether there is a correlation between atmospheric metallicity and fundamental planetary properties such as mass and semi-major axis (Figure~\ref{metallicity_mass_correlation}). 
A strong correlation is found in Solar System gas giants and there have been indications of such a correlation in exoplanet data, although the uncertainties in the atmospheric metallicity of exoplanets are still high (Figure~\ref{metallicity_mass_correlation}). Measurements of the Methane absorption line serve as our primary proxy for the atmospheric metallicity, allowing for direct comparison with the Solar System's outer planets \cite{kreidberg14}. Our quantitative objective is to achieve a measurement in the
correlation between planet mass and atmospheric metallicity with at least 3$-\sigma$ significance.

\begin{figure}\begin{center}
    \includegraphics[width=4in]{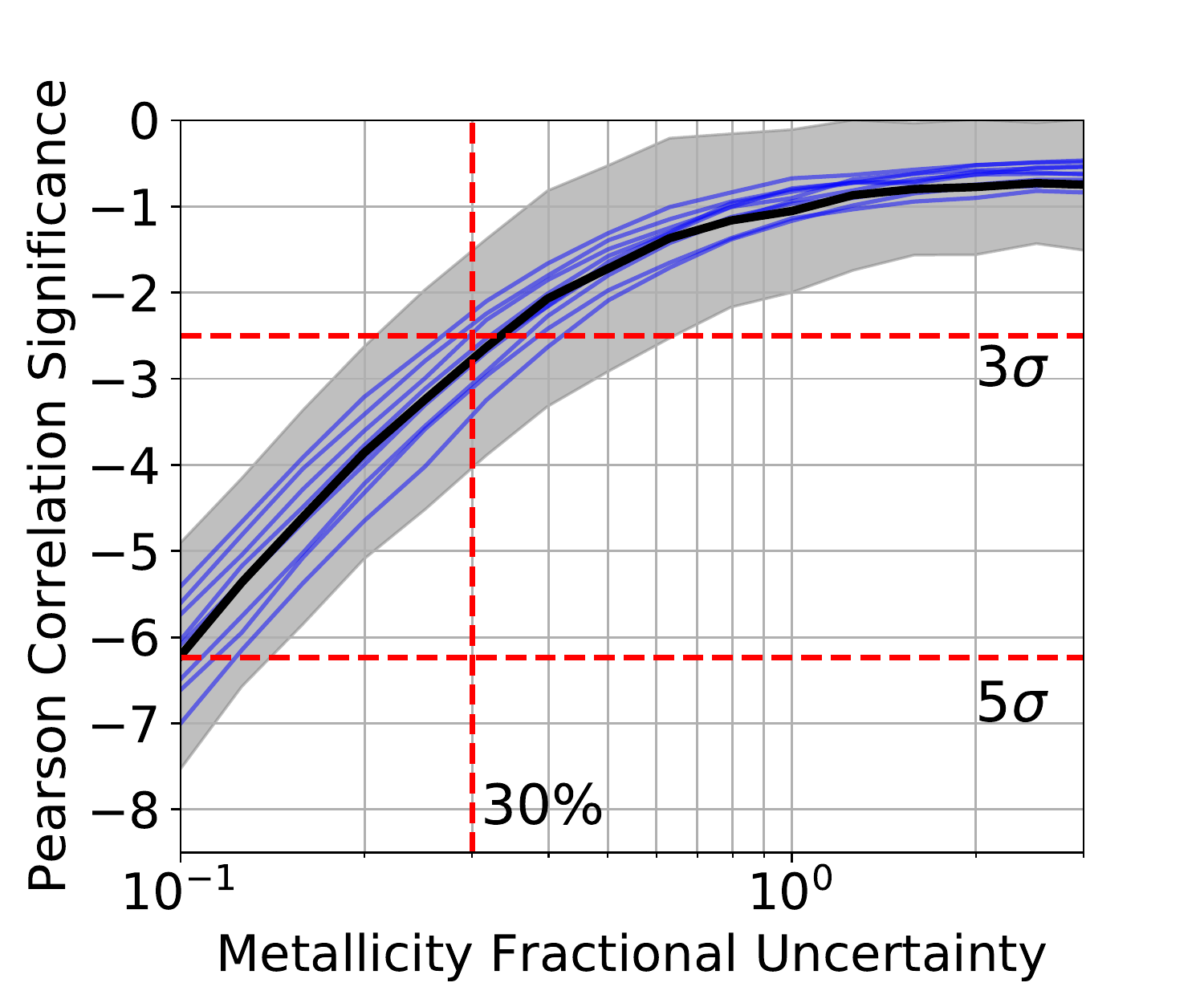}
  \end{center}\caption{
    Our ability to fit the mass-metallicity relationship among known gas-giant planets depends on the accuracy of the individual metallicity measurements. For several sets of 10 random-selected targets, the statistical significance of the Pearson correlation coefficient is estimated over a range of metallicity fractional uncertainties (purple lines). The median result is in black, with a gray band as the full range of uncertainties in the Pearson correlation significance. We find that a 30\% metallicity uncertainty is sufficient to assess the mass-metallicity relationship. The gray band shows the uncertainty on the Pearson correlation significance corresponding to the single instance represented by the thick black line.
  }\label{pearsonCorrelation}
\end{figure}

To determine if such a correlation is present in a population of gas giant exoplanets orbiting different stars, we need to establish how many are needed and with what level of uncertainty in metallicity. There are currently 20 gas giant exoplanets with orbital angular separations accessible to the Starshade Rendezvous Probe (from 0$\farcs$13 to 3$\farcs$2; Table~\ref{targetList-RV}).
Since it is unrealistic to expect all of them will be at an orbital phase favorable for spectral measurements, we looked at randomly sampled subsets that might be available. We find that if a subset of 10 stars is available with a 30\% metallicity fraction uncertainty, then it is possible to discern a mass-metallicity correlation with 3-$\sigma$ statistical significance (Figure~\ref{pearsonCorrelation}). The 30\% metallicity fractional uncertainty can be achieved with spectral SNR$>$15 measurements in one or two bands from $\sim$600 to $\sim$800 nm \cite{lupu16, damiano20}.

\input{targetList_RVplanets_new}

The mission requirements based on the driving investigation (detecting Earth-like planets) enable measurement of known giant-planet metallicities.
Integration times depend on the illumination phase angle during observation; while the orbital phase for radial-velocity-detected planets is known, the orbital inclination is uncertain. Table~\ref{targetList-RV} shows the integration times required to achieve spectral SNR$>$15 for a range of illumination phase angles. 
A geometric albedo of 0.3 is assumed, with radius constrained by the mass-size relation of \citenum{chen17} but conservatively capped at 1 Jupiter radius.
This albedo is a conservative choice, well below Jupiter's actual value of 0.5
\cite{karkoschka94,cahoy10},
but consistent with models of Jupiter-mass planets located closer to the Sun \cite{cahoy10, lupu16}. Note that the expected integration times have changed somewhat from what was reported in the Probe study report~\cite{seager19} due to changing the geometric albedo from 0.5 to 0.3 and the shift from the integral field spectrometer to the slit prism spectrometer, which changed the current best estimate of the end-to-end efficiency from 2.5\% to 3.4\%. 

Integration times are on the order of several days for the majority of targets at favorable illumination angles.  Ten spectra can be obtained with a total of 50 days of integration time allocated among the most favorable targets. Note that by the time starshade begins operations this target list is expected to have grown, providing even more flexibility in scheduling.
The observing windows for current targets are plotted in Figure~\ref{metallicity_windows} showing whether a target is available for observation at any given time during operations. \wfirst-CGI observations, continued Doppler monitoring, and Gaia astrometry will constrain the brightness and orbital parameters prior to starshade operations, so these observations can be precisely planned for maximum planet visibility with no need for revisits. 

\subsection{Sensitivity to Exozodiacal Dust}

The dust surrounding Earth-like planets is small in mass but large in area,
making it generally much easier to observe that the planet itself. The integrated flux from Solar System's zodiacal dust, for example, is orders of magnitude brighter than the Earth. 
Imaging the dust disk distribution requires an integration time of 1 day (on average) or up to 4 days (maximum) to obtain a flux sensitivity of 0.1 zodi, enabling 5$-\sigma$ detection of disks as faint as 0.5 zodi. 
While disks this faint have never been observed (other than the Solar System, with 1 zodi), the median level inferred from a sample of nearby stars is 4.5 zodis \cite{ertel20}, suggesting that most, if not all, of
the systems with habitable zone dust will be detected. With a telescope imaging resolution of $0\farcs065$, the disks will be mapped at spatial resolutions of $\sim$0.2--0.5 AU.

With this sensitivity it may be possible to detect the influence of planets on the zodiacal dust disk structure.  \citenum{stark08} have shown that planets with $\gtrsim$4 Earth masses can introduce significant features on the dusk disk brightness distribution of moderately bright dusts disks ($\sim$6--10 zodis). 
The induced structure would be located in close proximity to the observed planet --
with both following the same orbital trajectory -- removing any ambiguity
in whether the disk structure is planet related.

\section{Observing Strategy}\label{strategy}

The science objectives are met with an observing program
that balances detection and characterization of new exoplanets with
the characterization of known giant planets.
The observing strategy is guided by the sensitivity toward
individual targets (\S\ref{sensitivity})
and fundamental limits on the targets' visibility.

\subsection{Earth-Like Planets}

Having already identified the best targets for detection of Earth-like planets for SRP given its constraints(\S\ref{earthTargets}), we now describe our strategy on how to schedule observations
to optimize the number of characterized planets.
The 8 most promising stars will be given priority for at least one visit.
The revisit strategy for these targets depends on the information gathered during each visit. 

\begin{figure}\begin{center}
     \includegraphics[height=4.5in,angle=90]{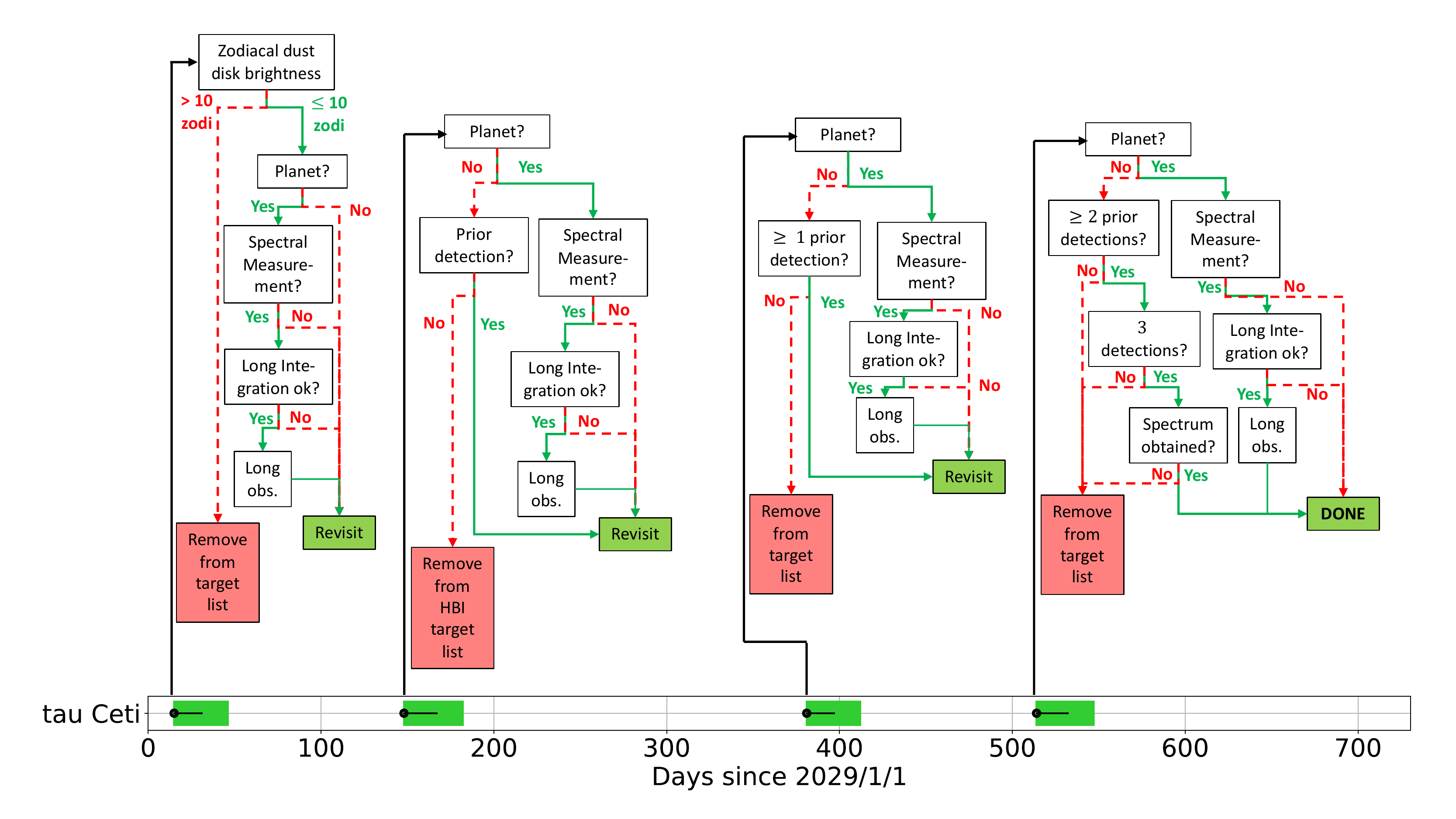}
  \end{center}
  \vspace{-0.4in}
  \caption{
    The decision tree for discovery of Earth-like planets
    adapts its observations as new information is gathered.
    Depending on 1) the amount of exozodiacal dust,
    2) the number of observations of a planet candidate, and
    3) whether long spectroscopic integration times are executed,
    a target may either be revisited or removed from the target list.
    The 2-year timeline on the bottom of the figure shows the windows available for observations (shown here with tau Ceti's observing windows).
  }\label{decisionTree}
\end{figure}

The decision tree used for observations is illustrated in Figure~\ref{decisionTree} and described in detail here. The decision tree assumes that the imaging observation is complete with $\sim$1 day of integration and the data is expected to be available for analysis within a couple of days. In the time between the observation and data retrieval the \wfirst\ telescope will be available for other observations while the starshade remains in position. The starshade science team will have fast analysis tools in hand to estimate the brightness of the exozodiacal dust disk and detect Earth-like exoplanet candidates. Based on the findings, the starshade team can either decide to initiate the cruise into position for observing the next target in the sequence or to initiate a long integration time observation for spectral characterization.

The first visit is a reconnaissance observation. The first check is whether the system has a exozodiacal dust disk above or below 10~zodi. While some of our target stars have existing upper limits on their exozodiacal dust disk brightness, based on precision nulling measurements of their dust's infrared emission, they are not sufficient to rule out deleterious levels of dust. 
Note that while we have used a fiducial value of 4.5~zodi in \S\ref{sensitivity} and \S\ref{results}, for the purposes of estimating background, in reality the exozodiacal dust brightness will be vary from target to target and we assume it will be unknown prior to the first observation. Here we are describing the tentative design reference mission for Starshade Rendezvous Probe mission. 
If the disk is brighter than 10~zodi, the target is removed from the habitability and biosignatures target list since it is not expected that an Earth-like exoplanet could be spectrally characterized against such a bright background as well as an increased risk of false-positive planet detections 
from planet-induced dust structures \cite{defrere12}.
If a target is removed, the observation plan will be updated with the next best target, which may or may not be visited at a later period depending on what is discovered for the target ensemble.

If the analysis finds that the exozodiacal dust disk is $\leq$10 zodi but no planet consistent with HZ orbit is found, the target is kept on the list for a revisit as a planet could still appear in subsequent observations.  If a planet consistent with a HZ orbit is found, 
then the imaging data will provide the brightness of the planet, which determines whether a spectral measurement with SNR $>$ 20 is achievable in the remainder of the observing time window, typically 25 days. If that is the case, a spectral observation will be initiated.

On the second visit to a target, if no planet consistent with an HZ orbit has been detected, then the target is removed and the observation plan is updated with the next best target in the list. If the planet is detected either for the first or second time, a decision is made based on the data to take a spectroscopic measurement as described above. Spectroscopic measurements are only required to be performed once, so if such a measurement was made in the first visit, then it will not be repeated.

On the third visit, if the planet consistent with a HZ orbit has only been detected once, no further visits will be planned since at least 3 observations in 4 planned visits are required. However, there is some flexibility in this decision since the third visit will be made in the second year of observations and the science team will have additional information on the exozodiacal dust disk brightness and population of planet candidates in the ensemble of targets already observed.
In the event that there are a small number of target systems left, then this system could have more visits planned. If there are a large number of relevant target systems still available, this revisit priority may fall lower than the priority of these other systems,recognizing revisit priority may evolve as knowledge is gained about each system.Spectroscopic measurements can be triggered based on the criteria discussed above.

In the event that the planet has only been observed twice before and/or no spectroscopic measurement has been made yet, a fourth visit will be needed to determining whether the planet's orbit is in the habitable zone. This will occur during the last $\sim$6~months of the mission so the prioritization of targets could be significantly affected by how many Earth-like exoplanet candidates have been found and their potential for a spectral measurements with SNR$\geq$20. 

\subsection{Known Gas-Giant Planets}

Unlike the habitability and biosignatures targets, it is expected that significantly more information about the known gas giants will be available. 
Although we have not assumed prior information on the targets obtained with the CGI in this study, it is possible that  \wfirst-CGI will have already observed these systems and directly imaged the planets before Starshade operations begin. 
This will determine how bright they are and whether or not spectral measurements with SNR$\geq$15 are viable. The top-ranked targets with their estimated integration time will be integrated into the observation plan. 

\subsection{Exozodiacal Dust}

No additional observations are required to observe the dusty debris
that is prevalent in planetary systems; it will be detected alongside
any observed planets.

For the aim of characterizing the influence of planets in the dust disk distribution, target stars with bright exozodi (6–-10 zodi) that show clumps with $\geq$10\% excess brightness may be revisited up to three more times.
Identifying Earth-like planets takes priority, but in the event that exozodiacal dust is too bright in most stars and systems with the characteristics described above exist, these observations will be executed.
These observations will track the motion of dust disk clumps to test whether their orbits are Keplerian, indicative of an associated planet.
Provided a system within this exozodi range with a $\geq 4 \MEarth$ planet is found, this measurement will provide a means to probe planetary systems in stars with high levels of exozodiacal dust in their habitable zones \cite{stark08}.

\subsection{Scheduling}

The visit strategy needs to be dynamic since it will be modified with each observation of habitability targets while at the same time ensuring that a subset of $10$ known gas giants can be spectrally characterized with retargets that optimize the fuel usage. Since there is a significant amount of uncertainty associated with the distribution of exozodiacal dust disks and the frequency of occurrence of Earth-like exoplanets, this requires a fairly sophisticated Monte Carlo simulation that demonstrates that the decision tree is adaptable to the full range of possibilities. This will be the subject of a future study. 

\begin{figure}\begin{center}
    \includegraphics[width=6.5in]{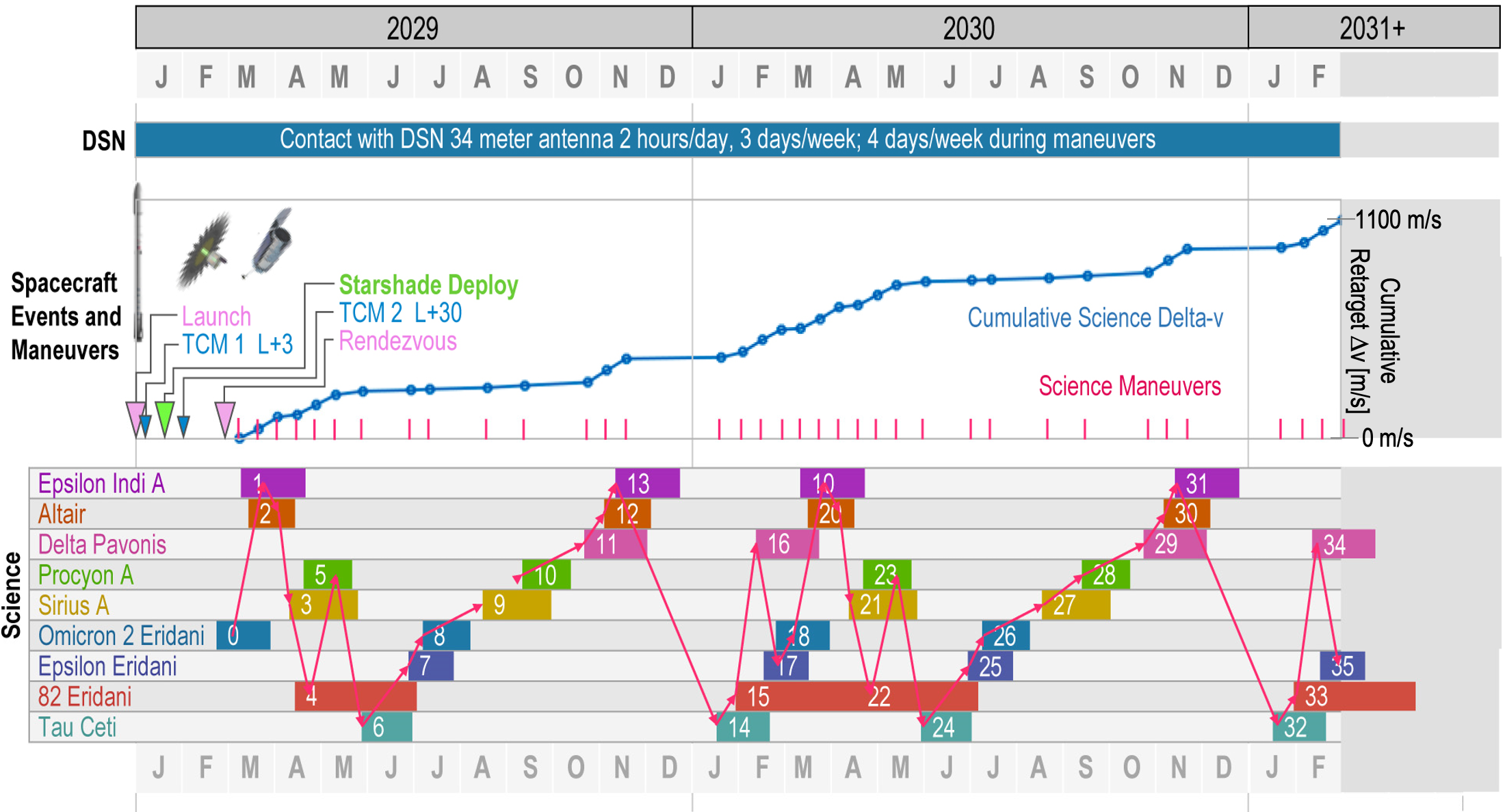}
  \end{center}\caption{
    An example of a retargeting strategy that bounds the fuel usage for the Starshade Rendezvous Probe mission. We have allocated 36 target maneuvers to accomplish the objectives. In this case, we have chosen a stressing case where the 36 retarget maneuvers are applied to 9 targets, each visited four times. While this is not representative of the mission, which would visit 10 habitable exoplanet targets with a subset of them requiring revisits in addition to the known gas giants, the reason this is a stressing case is that visiting a target twice a year on different availability windows requires more significant burns than having to visit a target only once a year or just once during the mission as will be that case for the known gas giants and habitability targets with large exozodiacal dust disk backgrounds. The top panel show the main spacecraft events and maneuvers along with the cumulative $\Delta$v estimates for each target as a function of time, labeled in year and month of year at the top of the figure. The vertical magenta tick marks show the time at which each target is visited. The bottom panel shows the targets chosen with their observing availability windows shown as horizontal bars. Each bar has numerical labels representing the order of the visits. 
    The timing of translational retargeting slews (red lines) may take a couple days up to two weeks depending on the angular separation between targets.
    The targets are arranged vertically in order of right ascension with the observing windows for each target shown as horizontal colored bars.
    The observation days are chosen to be at the beginning of the window,
    to allow time for follow-up spectroscopy within the same window.
  }\label{timeline}
\end{figure}

In the Starshade Probe Study report, the delta-v allocated to retargeting was 1100~m/s
\cite{seager19}.
An additional 300~m/s of delta-v is allocated for stationkeeping.
The allocation for large slews was estimated to be sufficient for 
36 retargeting maneuvers 
using a limiting scenario where 9 targets were visited 4 times each (Figure~\ref{timeline}). 
For retargeting within other starshade mission concepts see \citenum{stark16,soto19}.
While our objective is to visit 10 habitability and biosignatures targets, some with multiple revisits, and 10 known gas giants only once, the extreme scenario bounds the fuel usage because it assumes the need to retarget in situations that are not necessarily the most fuel efficient. The gas giants, for example, can be visited when they are close to the path needed for other targets resulting in significantly smaller fuel burns. 

The main driver for delta-v is the need to visit targets 4 times in a period of 2 years. If the mission duration could be made longer, the retargeting strategy could follow the natural right ascension progression of observing windows as the Earth rotates around the Sun, which can result in highly efficient fuel burns. With a two year window, there are cases where fairly aggressive burns are required to catch targets of interest when their observing windows are available. 

\section{Expected Performance}\label{results}

The expected scientific yield depends not only on the assumed mission parameters,
but also on exoplanet demographics.
While the frequency of gas-giant planets is relatively well known,
the probability that each of our target stars will have a habitable-zone
Earth-size planet ($\nEarth$) is not well constrained.
NASA's ExoPAG Study Analysis Group (SAG-13)
performed a meta-analysis of several published fits to Kepler survey results,
producing a planet frequency formula as a function of planet size and location
\cite{belikov17}.
Combining this formula with our adopted habitable zone and Earth-like planet radius definitions (Table~\ref{planetParamTable}),
we calculate an Earth-like planet frequency of $\nEarth = 0.24^{+0.3}_{-0.1}$.
Note that this definition of Earth-like planets and corresponding frequency 
matches for consistency Ref.~\citenum{gaudi20};
for possible alternative calculations of planet occurrence rates, 
see e.g.\ Refs. \citenum{hsu19,bryson20}.

\begin{figure}\begin{center}
    \includegraphics[width=4in]{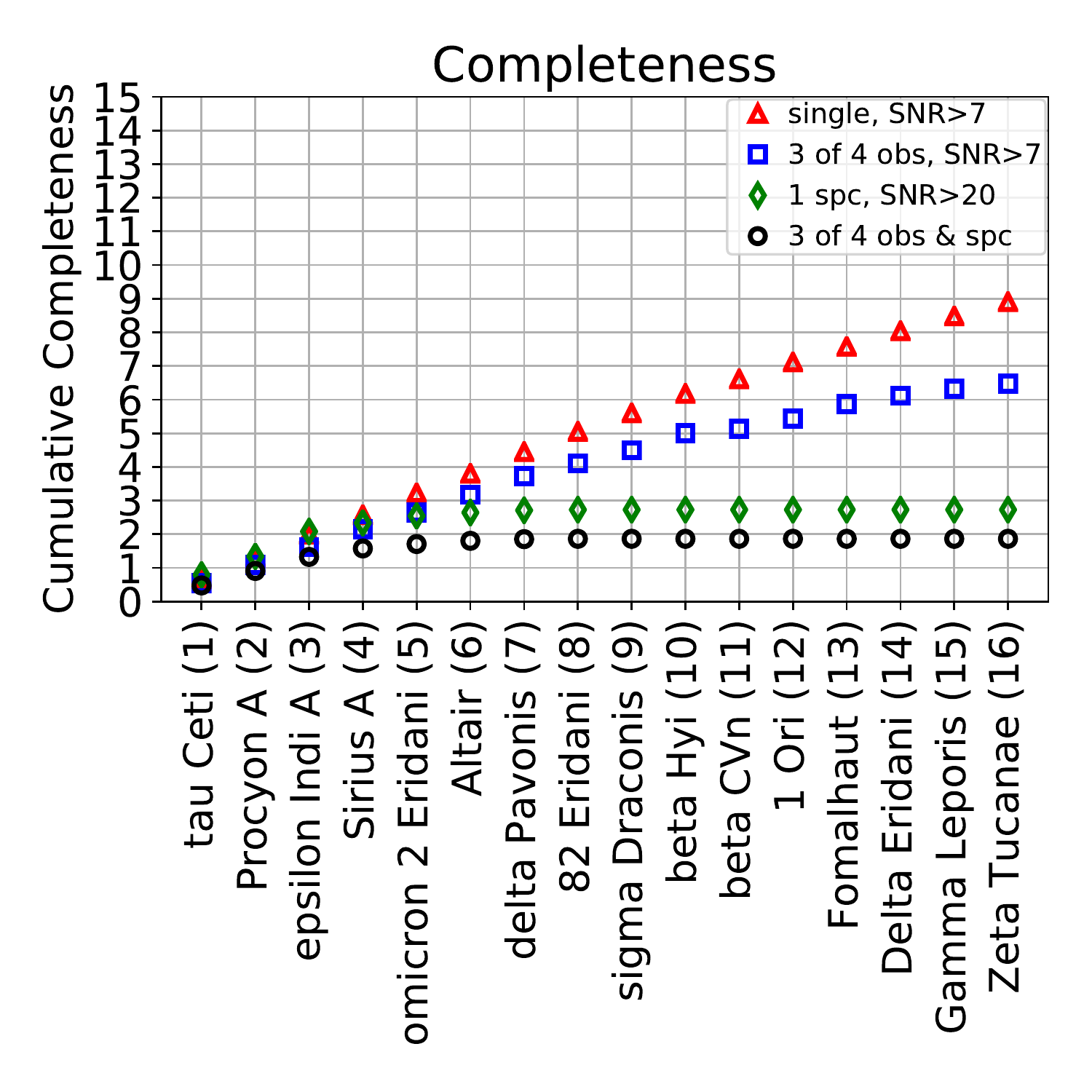}
  \end{center}\caption{
  Cumulative completeness for habitability and biosignatures targets (based on Figure~\ref{completeness}).
  }\label{fig:cumu_completeness}
\end{figure}

The cumulative completeness for habitability and biosignatures targets is shown in Figure~\ref{fig:cumu_completeness}. The expected number of Earth-like exoplanets is derived from multiplying the completeness by the occurrence rate $\eta_\oplus$. For a nominal observing program targeting at least 10 nearby stars
the expected number of detected Earth-like exoplanets is $1.5^{+1.9}_{-0.6}$. However, this is only for single visit detection. The number with orbits constrained to the habitable zone is $1.2^{+1.5}_{-0.5}$. Note that the cumulative completeness for spectral measurements flattens after the 5th target. The expected number of Earth-like exoplanets with spectral characterization is $0.65^{+0.82}_{-0.27}$. The expected number with both orbits constrained and a successful spectral measurement is $0.45^{+0.56}_{-0.19}$.
This number is lower than the expectation of $\sim$4 derived by Ref. \citenum{stark16} for the same telescope size and starshade launch mass,
primarily because \wfirst's actual end-to-end efficiency is lower than previously assumed, resulting
in significantly longer integration times for high-quality spectra.

\begin{figure}\begin{center}
    \includegraphics[width=4in]{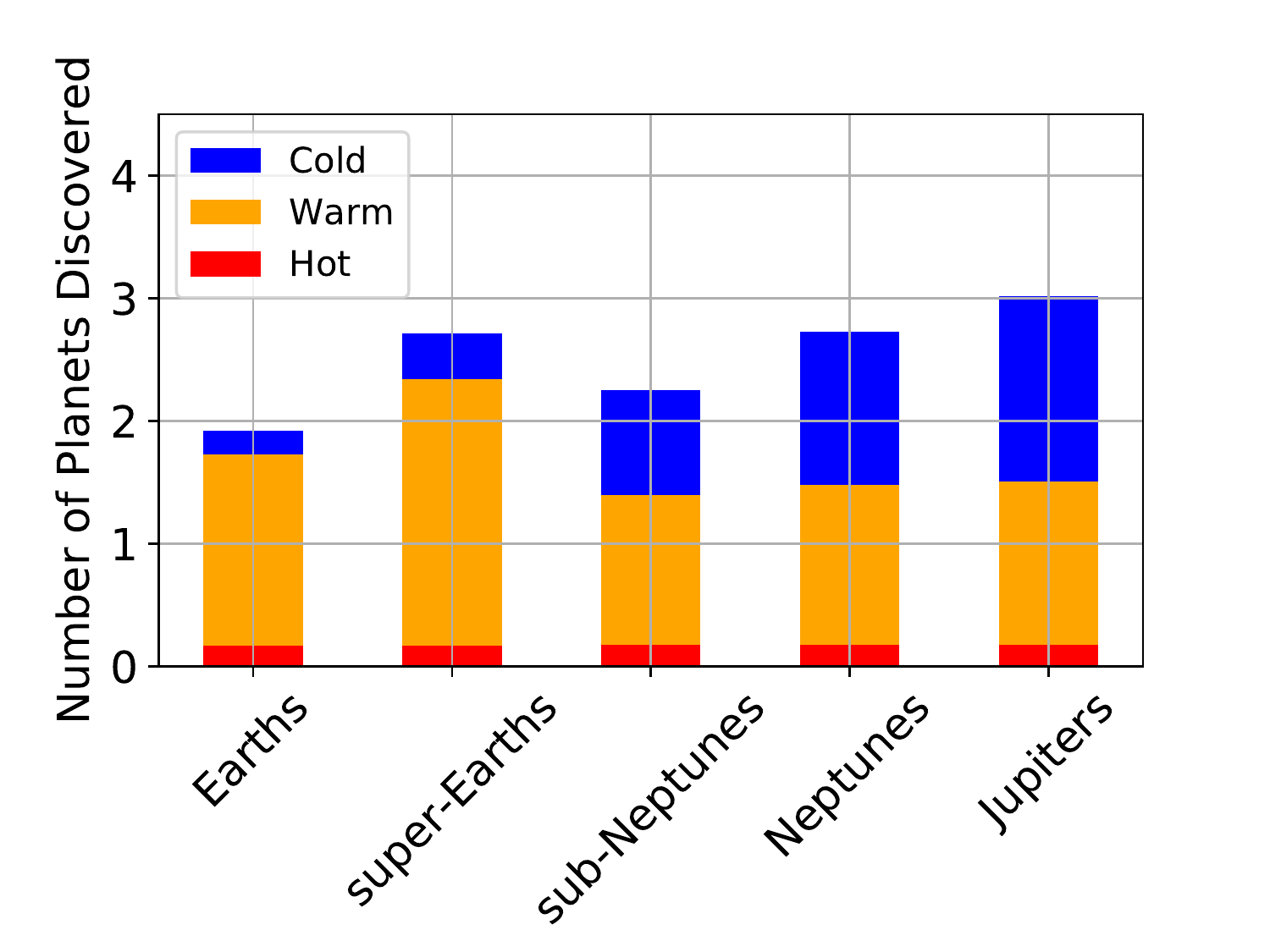}
  \end{center}\caption{
    Planet yield as a function of planet type and approximate temperature.
    The yield is obtained based on the single-visit completeness assuming detection with SNR $\geq$ 7.
    All observations assume a zodiacal dust disk brightness of 4.5 zodi.
    The bar chart assumes that the 16 targets for the habitability and biosignature gases investigation are visited at least once. The number of planets scales with the number of targets visited according the the single-visit completeness curve in Figure~\ref{fig:cumu_completeness}.
  }\label{yields}
\end{figure}

It is important to note that \wfirst\ observations with a starshade will be sensitive to a wide variety of planets.
Figure~\ref{yields}
shows the expected number of planets discovered by imaging (SNR$>$7), 
some of which are bright enough to obtain follow-up spectra.
This threshold SNR value is relaxed compared to Earth-like planets because giant exoplanets tend to have more pronounced spectral features.
The planet properties and frequency of occurrence are the same as the Exo-S report \cite{seager15} with the modification that the occurrence rate for warm Earth's and super-Earths was raised to 0.24.
These results indicate that $\sim$12 new planets will be discovered in the nearest sunlike stars providing additional information on their planetary system architectures.

\section{Conclusions}\label{summary}

We have presented the modeling, observing approach, and expected performance to meet the objectives of the Starshade Rendezvous Probe study \cite{seager19}. The Starshade Rendezvous Probe concept has the capability to deliver first-of-a-kind exoplanet direct imaging and spectroscopy results in the next decade. A deep-dive investigation will provide the first examination of planetary systems around our nearest sunlike stars, including their habitable zones, giant exoplanets, and warm dust disks, opening a new frontier. 

The Starshade Rendezvous Probe concept is capable of discovering Earth-size planets in the habitable zones of nearby stars using the relatively moderate aperture \wfirst\ space telescope. By initially characterizing the sensitivity to each individual target, we have found that while the SRP has the sensitivity to detect Earth-like exoplanets and constrain their orbits to the habitable zone, its primary limitation is the sensitivity to spectral measurements. 
The main means for improving this is to increase the aperture of the telescope, as would be done with HabEx, since this has the dual benefit of increasing the photon rate and better resolving the exozodiacal background. 
It is worth noting there are large uncertainties in the occurrence rates of Earth-like exoplanets and the distribution of zodiacal dust disk brightness, which could result in a increased discovery potential of the SRP if nature behaves favorably. 
The SRP is the only observatory that would have the capability to detect Earth-like exoplanets  within the next decade.

Observations of known planets with the SRP could determine whether the atmospheric metallicity and mass of known giant exoplanets follows the correlation observed in our own Solar System, testing whether there is a trend in planetary formation.
Meeting these objectives will begin to answer the driving questions of whether Earth is unique and how the Solar System compares to the planetary systems orbiting our nearest sunlike stars. The SRP is well equipped to meet this objective with an expected increase in the number of known gas giants with radial velocity measurements as well as observations with CGI prior to the SRP operations.

The SRP will obtain measurements of the exozodiacal dust of nearby sunlike stars with unprecedented sensitivity. This information is key to the future of direct imaging observatories since the dust brightness distribution is not known well enough to pin down the level of background light expected for planet detection and characterization. 
The sensitivity of the SRP to dust disks provides additional scientific opportunities to investigate the influence of planets on the dust disk morphology provided such systems are found.    

The main challenge of observing with starshades -- retargeting with constrained time windows in a relatively short mission duration -- has been addressed with a decision tree that can accommodate the large degree of uncertainty associated with searching for Earth-like exoplanets. The observing plan adapts to new information as the targets are observed multiple times with predetermined criteria for deciding whether to revisit targets or take spectral measurements. We have estimated 36 retargeting maneuvers are necessary to meet the science objectives and we have bounded the amount of delta-v needed with a stress case scenario. The driving use of fuel is the occurrence of large angle retargeting maneuvers needed to observe some targets multiple times within the two-year duration of the mission. 
Depending on the realization of Earth-like exoplanet occurrence and exozodiacal dust, the SRP mission could have enough fuel for an extended mission to visit more targets. 

\acknowledgments
Part of this work was carried out at the Jet Propulsion Laboratory, California Institute of Technology, under a contract with the National Aeronautics and Space Administration. \copyright 2020. All rights reserved.
This research has made use of 1) the NASA Exoplanet Archive, which is operated by the California Institute of Technology, under contract with the National Aeronautics and Space Administration under the Exoplanet Exploration Program and 2) the SIMBAD database, operated at CDS, Strasbourg, France. 


\bibliographystyle{spiejour}
\bibliography{refs}







\end{document}

%% file: missionParams.tex
\begin{deluxetable}{l|c}
\tablecaption{Mission Constraints
\label{missionParamTable}}
\tablehead{Parameter & Expected Performance}
\startdata
Starshade nominal mission lifetime         & 2 years \\
Telescope primary mirror & 2.4 m \\
\hline
Solar exclusion angle (min)  & 54\degree \\
Solar exclusion angle (max)  & 83\degree \\
\hline
Detector bandpass & 400 -- 1000 nm \\
Imaging resolution & 65 mas at 750 nm \\
Imaging end-to-end efficiency & 0.035$^a$ \\
Imaging Field of view (FOV) & 4.5\arcsec (radial) \\
\hline
\hline
\enddata
\tablenotetext{a}{see Table \ref{throughputTable} for details}
\end{deluxetable}

%% file: throughputTable.tex
\begin{deluxetable}{l|cc}
\tablecaption{Telescope efficiency ($\epsilon$)
\label{throughputTable}}
\tablehead{Contribution & \multicolumn{2}{c}{Best Estimate} }
\startdata
Geometric obscuration of the WFIRST pupil & \multicolumn{2}{c}{0.82} \\
Reflection losses in the telescope optics & \multicolumn{2}{c}{0.81} \\
Reflection \& transmission losses (excluding coronagraph masks) &
{0.60 (P)} & {0.58 (S)}  \\
Starshade dichroic beam splitter & \multicolumn{2}{c}{0.90} \\
Detector effective QE (at end of life) & \multicolumn{2}{c}{0.285} \\
Core throughput losses due to diffraction from WFIRST pupil &
\multicolumn{2}{c}{0.34} \\
\hline
Total & 0.035 (P) & 0.034 (S) \\
\hline
\enddata
\tablecomments{P = Photometry; S = Spectroscopy}
\end{deluxetable}

%% file: missionPossibilities.tex
\begin{deluxetable}{lcc}
\tablecaption{Assumed Mission Parameters
\label{missionPossibilitiesTable}}
\tablehead{Parameter & Assumed Performance}
\startdata
Time allocation~\cite{seager19}          & 136 days \\
Inner working angle (IWA) & 100 mas \\
Instrument contrast & $4 \times 10^{-11}$ \\
Delta-V for retargeting & 1,100 m/s \\
\hline
\enddata
\end{deluxetable}

%% file: planetParams.tex
\begin{deluxetable}{l|c}
\tablecaption{Assumed Planet and Dust Properties
\label{planetParamTable}}
\tablehead{Parameter & Value (or [Range])}
\startdata
Earth-like planet geometric albedo$^a$ ($A_G$) & 0.2 \\
Earth-like planet radius ($r_{p}$)  &  $[0.8 (R_{\rm pl}/{\rm AU})^{1/2} (L_\star/L_\odot)^{-1/4}$, 1.4] $\REarth$ \\
Habitable Zone ($R_{\rm pl}$) & $[0.95, 1.67] (L_{\star}/L_\odot)^{1/2}$ AU \\
\hline
Gas-giant planet geometric albedo ($A_G$)  & 0.3 \\
Gas-giant planet radius$^b$ ($r_{p}$) & Ref.\ \citenum{chen17} ($\RJup$ max)  \\
\hline
Zodiacal dust brightness ($dF_z/d\Omega$) & Ref.\ \citenum{leinert98}  \\
Exozodi dust brightness$^c$ ($dF_{ez}/d\Omega$) &  4.5 zodi \\
\hline
\enddata
\tablenotetext{a}{For the assumed isotropic scattering, this geometric albedo
  is equivalent to 0.3 spherical albedo.}
\tablenotetext{b}{While the radius depends on the mass of the gas giant planet, 
  we set a conservative upper limit of $\RJup$.}
\tablenotetext{c}{The unit of 1 zodi is equivalent to 22 mag/arcsec$^2$.}
\end{deluxetable}

%% file: targetList.tex
\setlength{\tabcolsep}{4pt} 
\begin{deluxetable}{l|cccccc|cccc}
\tablecaption{Nearby stars targeted for Earth-like planets
\label{targetList-HZ}}
\tablehead{Star & Distance & $V$ & $\Lstar$ & $T_{\rm eff}$ & $M_{\star}$ & Spectral & 
   \multicolumn{4}{c}{Completeness} \\
Name & (pc) & (mag) & ($\LSun$) & (K) & ($M_\odot$) & Type &
   Single-visit & Orbit & Spectral & Overall }
\startdata
tau Ceti$^{bc}$& 3.7 & 3.5  & 0.52  & 5283 & 0.80 & G8.5V  & 0.67 & 0.55 & 0.79 & 0.48 \\
Procyon$^a$    & 3.5 & 0.4  & 7.1   & 6543 & 1.49 & F5IV-V & 0.65 & 0.54 & 0.55 & 0.43 \\
eps Ind$^{ac}$& 3.6 & 4.7  & 0.23  & 4683 & 0.68 & K4V     & 0.67 & 0.52 & 0.74 & 0.42 \\
Sirius$^a$     & 2.6 &$-$1.4& 30.5  & 9580 & 2.40 & A1.0V  & 0.58 & 0.52 & 0.25 & 0.25 \\
\hline
omi 2 Eri$^c$  & 5.0 & 4.4 & 0.42  & 5151 & 0.81 & K0.5V   & 0.65 & 0.51 & 0.21 & 0.13 \\
Altair         & 5.1 & 0.8 & 10.7  & 7800 & 1.83 & A7IV-V  & 0.58 & 0.52 & 0.10 & 0.09 \\
del Pav        & 6.1 & 3.5 & 1.3   & 5590 & 0.99 & G8.0IV  & 0.64 & 0.55 & 0.07 & 0.05 \\
82 Eri$^c$     & 6.0 & 4.3 & 0.69  & 5401 & 0.85 & G8.0V   & 0.60 & 0.38 & 0.02 & 0.01 \\
\hline
sig Dra        & 5.8 & 4.7 & 0.44  & 5246 & 0.80 & G9.0V   & 0.55 & 0.39 & 0.00 & 0.00 \\
bet Hyi        & 7.5 & 2.8 & 3.7   & 5873 & 1.14 & G1IV    & 0.58 & 0.51 & 0.00 & 0.00 \\
bet CVn$^a$    & 8.4 & 4.2 & 1.3   & 5930 & 1.03 & G0V     & 0.43 & 0.13 & 0.00 & 0.00 \\
1 Ori          & 8.1 & 3.2 & 3.0   & 6424 & 1.24 & F6V     & 0.50 & 0.30 & 0.00 & 0.00 \\
\hline
Fomalhaut$^{ab}$&7.7 & 1.2 & 16.5 & 8399 & 2.05 & A3V      & 0.46 & 0.43 & 0.00 & 0.00 \\
del Eri        & 9.0 & 3.5 & 3.4  & 5095 & 1.19 & K0IV     & 0.46 & 0.25 & 0.00 & 0.00 \\
gam Lep        & 8.9 & 3.6 & 2.5  & 6372 & 1.27 & F7V      & 0.44 & 0.21 & 0.00 & 0.00 \\
zet Tuc        & 8.6 & 4.2 & 1.3  & 5948 & 1.01 & G0V      & 0.42 & 0.14 & 0.00 & 0.00 \\
\hline 
\enddata
\tablenotetext{a}{Binary (see Figure~\ref{binaryTargets})}
\tablenotetext{b}{Known debris disk}
\tablenotetext{c}{Known to have planet(s)}
\end{deluxetable}

%% file: targetList_RVplanets_tauCeti.tex
\begin{deluxetable}{l|cccccc|cccc|ccc}
\rotate
\tablecaption{Known super-Earth exoplanet target list
  \label{targetList-tauCeti}}
\tablehead{Planet & distance & $V$ & $\Lstar$ & $T_{\rm eff}$ & $M_{\star}$ & Spectral &
  $M_p$ & \multicolumn{2}{c}{$a_p$} & $F_p/F_\star$$^a$ &
  \multicolumn{3}{c}{Integration time (days)$^b$} \\
  Name & (pc) & (mag) & ($\LSun$) & (K) & ($\MSun$) & Type &
  ($M_{\oplus}$) & AU & (mas) & ($\times 10^{-9}$) &
    $\beta$=45\degree & $\beta$=90\degree & $\beta$=135\degree}
\startdata
tau Ceti e &   3.7 & 3.49 &  0.5 & 5283 &  0.8 & G8.5V & 3.9 & 0.54 & 147 & 1.86 
 & 0.14 &     0.44   &      9.3    \\
tau Ceti f &   3.7 & 3.49 &  0.5 & 5283 &  0.8 & G8.5V & 3.9 & 1.33 & 365 & 0.30
 &   1.9      &        8.4 &      -     \\
\enddata
\tablenotetext{a}{Planet-star flux ratio for a half-illuminated planet ($\beta$=90\degree).}
\tablenotetext{b}{Integration times to reach SNR=20 for an R=50 spectra,
  over a range of illumination phase angles $\beta$.}
\tablecomments{Dashes indicate integration times in excess of 25 days.}
\end{deluxetable}

%% file: targetList_RVplanets_new.tex
\begin{deluxetable}{l|cccccc|cccc|ccc}
\rotate
\tablecaption{Known gas-giant exoplanet target list
  \label{targetList-RV}}
\tablehead{Planet & distance & $V$ & $\Lstar$ & $T_{\rm eff}$ & $M_{\star}$ & Spectral &
  $M_p$ & \multicolumn{2}{c}{$a_p$} & $F_p/F_\star$$^a$ &
  \multicolumn{3}{c}{Integration time (days)$^b$} \\
  Name & (pc) & (mag) & ($\LSun$) & (K) & ($\MSun$) & Type &
  ($M_{\rm Jup}$) & AU & (mas) & ($\times 10^{-9}$) &
    $\beta$=45\degree & 90\degree & 135\degree}
\startdata
bet Gem b &  10.4 & 1.16 & 40.9 & 4850 &  2.6 & K0IIIvar & 2.30 & 1.64 &  158 & 10.9 & $\leq 0.1$ & $\leq 0.1$ & $\leq 0.1$ \\
gam Cep b $^c$ &  14.1 & 3.21 & 11.8 & 4761 &  1.9 & K1IV & 1.85 & 2.05 &  145 & 7.13 & $\leq 0.1$ & $\leq 0.1$ & 0.5 \\
ups And d &  13.5 & 4.09 &  3.6 & 6213 &  1.3 & F8V & 4.13 & 2.51 &  186 & 4.43 & $\leq 0.1$ & 0.3 &  7 \\
eps Eri b &   3.2 & 3.71 &  0.4 & 5146 &  0.9 & K2.0V & 1.55 & 3.39 & 1055 & 2.65 & $\leq 0.1$ & 0.3 &  7 \\
47 UMa b &  14.1 & 5.03 &  1.7 & 5882 &  1.1 & G0V & 2.53 & 2.10 &  149 & 6.57 & $\leq 0.1$&  0.6 &  16 \\
\hline
47 UMa c &  14.1 & 5.03 &  1.7 & 5882 &  1.1 & G0V & 0.54 & 3.60 &  255 & 2.59 & 0.8 &  3.5 &   - \\
HD 192310 c $^c$ &   8.9 & 5.72 &  0.4 & 5080 &  0.8 & K3V & 0.08 & 1.18 &  132 & 3.29 & 0.6 &  2 &  - \\
HD 219134 h $^c$ &   6.5 & 5.57 &  0.3 & 4835 &  0.8 & K3.0V & 0.34 & 3.11 &  475 & 2.79 & 0.7 &  3 &   - \\
HD 39091 b &  18.3 & 5.65 &  1.6 & 5950 &  1.1 & G1V & 10.02 & 3.10 &  169 & 2.67 &  1 &  6 &   - \\
HD 114613 b &  20.7 & 4.84 &  4.5 & 5782 &  1.2 & G3V & 0.36 & 5.34 &  258 & 1.00 &  2 &   10 &  -  \\
\hline
HD 190360 b &  15.9 & 5.73 &  1.2 & 5552 &  1.0 & G6IV & 1.54 & 3.97 &  250 & 1.93 &  3 &   15 &  -  \\
HD 160691 c &  15.5 & 5.12 &  2.0 & 5784 &  1.1 & G3IV/V & 1.81 & 5.24 &  337 & 1.09 &  3 &   15 &  -  \\
14 Her b &  17.6 & 6.61 &  0.7 & 5388 &  1.1 & K0V & 4.66 & 2.93 &  166 & 3.20 &  4 &   19 &  -  \\
55 Cnc d &  12.3 & 5.96 &  0.7 & 5235 &  1.0 & G8V & 3.88 & 5.50 &  445 & 0.92 &   11 &  -  &  -  \\
HD 154345 b &  18.6 & 6.76 &  0.7 & 5468 &  0.9 & G8V & 0.82 & 4.21 &  226 & 1.80 &  21  &  -  &  -  \\
\hline
HD 217107 c &  19.9 & 6.16 &  1.2 & 5704 &  1.1 & G8IV/V & 2.60 & 5.32 &  267 & 1.02 &   20 &  -  &  -  \\
HD 142 c &  25.7 & 5.70 &  3.0 & 6249 &  1.2 & F7V & 5.30 & 6.80 &  264 & 0.58 &  16 &  -  &  - \\
eps Ind A b $^c$ &   3.6 & 4.69 &  0.2 & 4683 &  0.7 & K4V & 3.25 & 11.55 & 3188 & 0.21 &   14 &  -  &  -  \\
GJ 229 b $^c$&   5.8 & 8.15 &  0.1 & 3709 &  0.5 & M1.5V & 0.03 & 0.90 &  156 & 1.66 &   12 &  -  &  -  \\
GJ 832 b &   4.9 & 8.67 &  0.0 & 3601 &  0.4 & M1.5V & 0.68 & 3.56 &  719 & 2.56 &  24  &  -  &  -  \\
\enddata
\tablenotetext{a}{Planet-star flux ratio for a half-illuminated planet ($\beta$=90\degree).}
\tablenotetext{b}{Integration times to reach SNR=15 for an R=50 spectra,
  over a range of illumination phase angles $\beta$.}
\tablenotetext{c}{Planet not listed in the SRP study report \cite{seager19}.}
\tablecomments{Dashes indicate integration times in excess of 25 days.}
\end{deluxetable}